\documentclass[sigconf, nonacm]{acmart}
\usepackage[inline]{enumitem}
\usepackage[nolist]{acronym}
\begin{acronym}

    \acro{AI}{Artificial Intelligence}

    \acro{HCI}{Human-Computer Interaction}

    \acro{IRR}{Inter-Rater Reliability}

    \acro{LLM}{Large Language Model}

    \acro{NLP}{Natural Language Processing}
    
    \acro{QDA}{Qualitative Data Analysis}

    \acro{SDM}{Shared Decision Making}
\end{acronym}

\usepackage{svg}
\usepackage{xspace}
\usepackage{subcaption}
\usepackage[capitalize,noabbrev]{cleveref}
\usepackage{makecell}

\definecolor{deleted}{RGB}{178,178,178}
\definecolor{inserted}{RGB}{14,130,27}

\newcommand{\xhdr}[1]{\par\addvspace{1mm}\noindent{\textbf{#1.}}}

\newsavebox{\summaryboxcontent}

\newenvironment{summarybox}{%
    \par\addvspace{\medskipamount}
    \begin{lrbox}{\summaryboxcontent}%
    \begin{minipage}{0.94\linewidth}%
}{%
    \end{minipage}%
    \end{lrbox}%
    \noindent
    \begingroup
    \setlength{\fboxsep}{8pt}%
    \colorbox{black!10}{\usebox{\summaryboxcontent}}%
    \endgroup
    \par\addvspace{\medskipamount}
}

\setlist[enumerate,1]{label=(\arabic*)}
\setlist[enumerate,2]{label=(\theenumi.\arabic*)}
\setlist[enumerate,3]{label=(\theenumii.\arabic*)}

\setlist[enumerate,1]{label=(\arabic*), ref=\arabic*}
\setlist[enumerate,2]{label=(\theenumi.\arabic*), ref=\theenumi.\arabic*}
\setlist[enumerate,3]{label=(\theenumii.\arabic*), ref=\theenumii.\arabic*}

\newcommand{\eg}{e.\,g.}

\newcommand{\etal}{et~al.\@\xspace}

\newenvironment{inumerate}{
    \begin{enumerate*}[label=(\alph*), itemjoin={{, }}, itemjoin*={{, and }}, after={{.}}]
}{
    \end{enumerate*}
}

\AtBeginDocument{%
  }

\begin{document}

\title[A Researcher Study on the Potential of AI Support for Qualitative Data Analysis]{From Assistance to Autonomy -- A Researcher Study on the Potential of AI Support for Qualitative Data Analysis}

\author{Elisabeth Kirsten}
\orcid{0009-0003-0680-8916}
\authornote{Work done while at Max Planck Institute for Security and Privacy.}
\affiliation{%
  \institution{Ruhr University Bochum}
  \city{Bochum}
  \country{Germany}
}
\email{elisabeth.kirsten@rub.de}

\author{Annalina Buckmann}
\orcid{0000-0002-7959-9743}
\affiliation{%
  \institution{Ruhr University Bochum}
  \city{Bochum}
  \country{Germany}
}
\email{annalina.buckmann@rub.de}

\author{Leona Lassak}
\orcid{0000-0001-8309-3211}
\affiliation{%
  \institution{Ruhr University Bochum}
  \city{Bochum}
  \country{Germany}
}
\email{leona.lassak@rub.de}

\author{Nele Borgert}
\orcid{0000-0002-0255-5822}
\affiliation{%
  \institution{University of Bern}
  \city{Bern}
  \country{Switzerland}
}
\email{nele.borgert@unibe.ch}

\author{Abraham Mhaidli}
\orcid{0000-0002-9519-245X}
\affiliation{%
  \institution{Max Planck Institute for Security and Privacy}
  \city{Bochum}
  \country{Germany}
}
\email{abraham.mhaidli@mpi-sp.org}

\author{Steffen Becker}
\orcid{0000-0001-7526-5597}
\affiliation{%
  \institution{Ruhr University Bochum}
  \city{Bochum}
  \country{Germany}
}
\additionalaffiliation{%
  \institution{Max Planck Institute for Security and Privacy}
  \city{Bochum}
  \country{Germany}
}
\email{steffen.becker@rub.de}

\renewcommand{\shortauthors}{Kirsten \etal}


\begin{abstract}
The advent of AI technologies, such as \aclp{LLM}, has introduced new possibilities for \acf{QDA}, offering both opportunities and challenges. 
To help navigate the responsible integration of \acs{AI} into \ac{QDA}, we conducted semi-structured interviews with 15 Human-Computer Interaction (\acs{HCI}) researchers experienced in \ac{QDA}. 
While our participants were open to \acs{AI} support in their \acs{QDA} workflows, they expressed concerns about data privacy, autonomy, and the quality of AI outputs.
In response, we developed a framework that spans from minimal to high \acs{AI} involvement, providing tangible scenarios for integrating \acs{AI} into \acs{QDA} practices while addressing researchers' needs and concerns. 
Aligned with real-life \acs{QDA} workflows, we identify potential for AI tools in areas such as data pre-processing, researcher onboarding, or conflict mediation.
Our framework aims to provoke further discussion on the development of \acs{AI}-supported \ac{QDA} and to help establish community standards for responsible Human-AI collaboration.
\end{abstract}


\maketitle

\section{Introduction}
\label{qdallm::sec::introduction}
\acf{AI} tools have grown increasingly popular, driven especially by the rapid rise of \acfp{LLM}. 
These models can generate coherent, contextually relevant text and perform tasks such as classification and summarization~\cite{weiEmergentAbilitiesLarge2022,bubeckSparksArtificialGeneral2023}.
They are increasingly adopted across domains, including research contexts.
One area where LLMs hold significant, yet debated potential is \acf{QDA}.
\ac{HCI} researchers are beginning to explore the use of \acp{LLM} throughout their research processes for tasks such as conducting \ac{QDA}~\cite{depaoliPerformingInductiveThematic2024}, generating synthetic data and substituting human participants~\cite{hamalainen2023evaluating,schmidt2024simulating,argyle2023out}, creative ideation~\cite{shaer2024ai,oppenlaender2023mapping}, literature reviews, and writing~\cite{kapania2025categorizing}.
Commercial tools have also started integrating \ac{AI} features (\eg, ATLAS.ti,\footnote{\url{https://atlasti.com/}} MAXQDA,\footnote{\url{https://www.maxqda.com/products/ai-assist}} and NVivo's Lumivero AI Assistant\footnote{\url{https://lumivero.com/product/nvivo-ai-assistant/}})~\cite{morganExploringUseArtificial2023}.

The rapid adoption of \ac{AI} tools, including LLM-based systems, begs the question: \textit{which steps of the \ac{QDA} workflow are suitable for AI support, and how do we make these determinations?}
These questions are particularly relevant for \ac{HCI}, where QDA plays a dual role: as a methodological tool, and as a site of collaboration, negotiation, and interpretive practice in interdisciplinary teams with different epistemic traditions~\cite{bjorn2015multiple, fiesler2019qualitative, mcdonald2019reliability}.
Although AI tools can be useful, they raise dilemmas about researcher autonomy~\cite{feustonPuttingToolsTheir2021} and risks (\eg, \acp{LLM} hallucinating wrong information)~\cite{Xiao2021}.
As a community, it is imperative to critically assess which tasks in the \ac{QDA} workflow are appropriate or inappropriate for AI integration, and to ensure its responsible use.
To better understand how AI tools can, should, or should not be integrated into QDA, we conducted semi-structured interviews with \ac{HCI} researchers about their \ac{QDA} workflows and opinions on \ac{AI} integration.
We found that while researchers' \ac{QDA} workflows share structural similarities, they differ significantly based on project specifics, highlighting the need for \ac{AI} tools that are highly customizable.
Researchers were generally open to \ac{AI} assistance when it enhanced their workflow but stressed the importance of ethical considerations, trust in \ac{AI} outputs, and maintaining control.

Our contributions are as follows. 
First, we provide an empirical account of researchers' \ac{QDA} practices, pain points, and challenges. 
Second, we present a framework for integrating AI into \ac{QDA} workflows, grounded in researchers' perspectives on the individual stages of \ac{QDA} and prior work, extending beyond the task of coding itself.
Based on our findings we outline tangible scenarios for the integration of AI, including AI as a productivity tool (\eg, taking on more technical tasks), as a mediator (\eg, facilitating consensus), or AI with human approval
(\eg, suggesting codebook components with validation by a researcher).
Finally, we discuss how this framework can serve as a reflective tool for researchers and teams to critically examine if, where, and how AI may be integrated into their QDA workflows.

\section{Background}
\label{qdallm::sec::background}
In this section, we lay out the methodological backgrounds of \acf{QDA} and trace the evolution of AI-assisted \ac{QDA} from basic computational methods to the growing role of \acfp{LLM}. We conclude by synthesizing our research questions.

Throughout the paper, we use \textit{AI} as a broad term for computational systems that support or automate parts of \ac{QDA}, including earlier NLP and machine learning methods as well as current generative systems.
We use \textit{\acp{LLM}} when referring specifically to language-model based systems, particularly where support involves open-ended natural-language generation.

\subsection{Qualitative Data Analysis in HCI Research}
\label{qdallm::sec::background::qda}
\ac{QDA} is a fundamental process in qualitative and mixed-method research within HCI, providing systematic techniques for interpreting non-numeric data such as interviews, field notes, videos, or documents~\cite{Booth.2012, mcdonald2019reliability}. 
Its primary goal is to juxtapose or integrate different sources to identify patterns, themes, and insights that deepen our understanding of the research subject. 
At its core, \ac{QDA} typically involves iterative cycles of data extraction, synthesis, and analysis~\cite{Booth.2012}. 

\ac{QDA} is shaped by various methodological frameworks that may overlap in scientific practice (see reporting of \ac{QDA} steps in \autoref{qdallm:results:rq1}) but are distinct in theoretical foundations, flexibility, and depth of analysis.
While \textit{Thematic Analysis}~\cite{braun2006using} is a theoretical framework that focuses on coding features relevant to the research question, \textit{Content Analysis}~\cite{wilkinson2000women} quantifies specific events or codes, enabling a blend with quantitative approaches. 
\textit{Grounded Theory}~\cite{glaser1967grounded} aims at theory generation through systematic coding directly from the data, and \textit{Framework Analysis}~\cite{barnett2009methods} applies existing frameworks, refining them throughout analysis, useful in applied research contexts that aim to build upon and extend existing knowledge.
In contrast, \textit{Narrative Analysis}~\cite{riessman2008narrative} emphasizes the structure and meaning of data within its narrative context.
Coding strategies can be \textit{deductive} (codes follow predefined categories), \textit{inductive} (patterns identified in the data), or hybrid, enabling researchers to choose an approach that aligns with their objectives.

The choice of coding strategy is not merely technical but reflects a researcher's underlying goals and questions, guiding the research direction and ensuring the chosen method supports the intended outcomes. 
Underlying these methods are different epistemologies, from more positivist ones aiming at recognition and discovery, to more interpretive ones, requiring in-depth analysis and centering researcher subjectivity and reflexivity~\cite{arthur2021social, greckhammer2022QualitativeResearch, chatzichristos2025qualitative, ashcroft2023reflexivity, soden2024evaluating}. 

Within HCI, projects commonly involve more than one re\-sear\-cher~\cite{mcdonald2019reliability, jiangSupportingSerendipityOpportunities2021}, making \ac{QDA} a fundamentally collaborative and social practice of negotiating, resolving disagreements, and making sense of ambiguity. 
This collaboration is often asynchronous and distributed, and supported by digital tools and software~\cite{wallace2017technologies}.
Researchers increasingly incorporate \ac{NLP} and \ac{LLM} methods into the QDA process -- thus, QDA fundamentally becomes \textit{computer-supported cooperative work}. 
At the same time, \ac{QDA} in \ac{HCI} is shaped by interdisciplinary diversity~\cite{bjorn2015multiple}.
Researchers bring different epistemological traditions and backgrounds~\cite{fiesler2019qualitative, mcdonald2019reliability, bjorn2015multiple}, which leads to ongoing debates about transparency, rigor, reliability, and measures thereof, such as \ac{IRR}~\cite{fiesler2019qualitative, mcdonald2019reliability, bjorn2015multiple, soden2024evaluating}. 
Yet, despite \ac{QDA}'s centrality to HCI, there has been no systematic documentation of how \ac{HCI} researchers actually conduct \ac{QDA} in practice, or how emerging technologies might reshape it.
The closest analysis we could find is the work by McDonald et al.\cite{mcdonald2019reliability} documenting how inter-rater reliability is used in \ac{HCI}. They find that reporting \ac{IRR} is rare and authors often use other methods to report on reliability. 
A more comprehensive overview of the whole QDA process, though, is still lacking. 
Our work addresses this gap by synthesizing and extending fragmented conversations into a coherent account of real-world \ac{QDA} workflows.

\subsection{From Computational Support to AI-Assisted Analysis}
\label{qdallm::sec::background::nlptollm}
Computer-assisted approaches have long supported \ac{QDA}, from supporting basic coding tasks, to experiments with automation.
Early tools applied basic \ac{NLP} techniques such as word frequency analysis, keyword matching, and pattern-based rules~\cite{stoneGeneralInquirerComputer1962, maratheSemiAutomatedCodingQualitative2018}.
While these methods introduced automation, they lacked the capacity to capture complex human interpretations.
Subsequent work introduced supervised machine learning for text classification, extending manual coding into larger datasets~\cite{rietzCodyAIBasedSystem2021}.
Unsupervised approaches such as topic modeling uncover latent themes without predefined categories~\cite{lennonDevelopingTestingAutomated2021}, and some studies reported surprising alignment with qualitative methods like grounded theory~\cite{baumer2017comparing,muller2016machine}.
Such combinations of QDA and machine learning can be especially useful for mixed-methods studies and for analyzing large-scale data, though they raise concerns about transparency and interpretability -- values central to qualitative research.
Recent advances in generative AI, particularly \acp{LLM} like ChatGPT, offer new opportunities.
These models are trained on massive text corpora and can generate, summarize, and classify language with a high degree of contextual awareness~\cite{weiEmergentAbilitiesLarge2022}. 
They encode vast amounts of diverse knowledge, making them potentially adaptable to various tasks within qualitative research.
Recent prototypes integrate LLMs at different points across the QDA workflow, from inductive coding to codebook development~\cite{banoAIHumanReasoning2023,taiExaminationUseLarge2024, xiaoSupportingQualitativeAnalysis2023, morganExploringUseArtificial2023, rasheedCanLargeLanguage2024, gauthier2022computational}. \acp{LLM} have shown remarkable performance on annotation tasks where little or no labeled data is provided~\cite{gilardiChatGPTOutperformsCrowdWorkers2023,tornbergChatGPT4OutperformsExperts2023} and have been proposed as \textit{collaborators} in human annotation teams~\cite{ziemsCanLargeLanguage2024}, indicating a shift from assistance to autonomy of computational methods supporting the cooperative work of QDA.

This evolution highlights not only the opportunities for scaling qualitative work, but also the risks it creates for preserving interpretive rigor and methodological integrity. 
These concerns are shared among HCI researchers~\cite{feustonPuttingToolsTheir2021, jiangSupportingSerendipityOpportunities2021, lubars2019ask}, expressing reservations about, or oppositions to, the integration of AI into QDA, or specifying specific use contexts that retain human agency. 
While these studies were conducted prior to the advent of LLMs, recent work identified a ``generational gap'', with early career researchers being more open, and more experienced researchers more reserved towards the integration of AI into QDA~\cite{chatzichristos2025qualitative}. 
\citet{chatzichristos2025qualitative} suggests that this skepticism among experienced researchers stems precisely from their deep understanding of reflexivity and contextual richness in qualitative inquiry, alongside potential structural factors such as resources and career stage.
This necessitates further engagement with shifting researcher attitudes towards AI-supported QDA.

\subsection{Potentials and Risks of AI in the QDA Workflow}
Recent studies show how AI can assist at multiple points in the QDA process. 
\citet{cuevas2025collecting} examine the usage of \acp{LLM} for conducting user studies and analyzing the collected data.
For labeling and annotating, LLM-based systems apply existing codes from a given codebook with notable consistency, though they often miss subtle contextual judgments~\cite{xiaoSupportingQualitativeAnalysis2023}.
\citet{lam2024concept} develop an LLM-based text analysis tool that derives interpretable concepts from unstructured text, inspired by traditional topic models.
For data exploration and inductive coding, \acp{LLM} can surface novel patterns and assist in the initial exploration of themes and codes, potentially enabling a deeper understanding.
However, while \acp{LLM} can replicate many primary themes identified by humans, they often miss more subtle and complex interpretations~\cite{morganExploringUseArtificial2023,zhangQualiGPTGPTEasyuse2023,torii2024expanding}.
\citet{hamiltonExploringUseAI2023} observed complementary patterns between human and ChatGPT-generated codes, with each identifying themes that the other did not.
\citet{baranyChatGPTEducationResearch2024} investigated the use of ChatGPT for codebook creation, finding hybrid approaches to be more reliable. They suggest LLMs can contribute to the improvement of codebooks, but stress that human involvement remains crucial. 
While combining interpretive and positivist methods can cause epistemic tensions and discussions among researchers~\cite{baumer2017comparing, muller2016machine, paulus2024minutes}, studies highlight mutual compatibility~\cite{baumer2017comparing, chatzichristos2025qualitative, gauthier2022computational, lam2024concept, muller2016machine}. AI also holds potential for mediating discussions and facilitating collaboration by highlighting coding discrepancies, and suggesting resolutions~\cite{gaoCoAIcoderExaminingEffectiveness2023, ganjiEaseCodeComplex2018,drouhardAeoniumVisualAnalytics2017}.

While these efforts illustrate potential, they also expose tensions.
Concerns persist around bias~\cite{bail2024can, tornberg2023use, ziemsCanLargeLanguage2024, anis2023efficient, owoahene2024review, banoAIHumanReasoning2023}, lack of reliability for non-English languages~\cite{heseltine2024large, suter2024using}, and the risk of treating AI as ``culturally agnostic'' despite its privileging of certain perspectives~\cite{paulus2024minutes}.
Such issues are particularly important when engaging minority groups or Global South contexts, where AI risks reproducing inequalities in representation and missing interpretative nuance~\cite{kumar2021braving, wacharamanotham2020transparency, rifat2023many, jang2023platform, okolo2022making}. 
This raises concerns regarding bias in interpretation, but also research ethics of fairness and inclusion~\cite{kempny2025potenziale}. 
While emerging work has begun to examine AI deployment in Global South contexts across various domains~\cite{okolo2022making, linxen2021weird, tahaei2024surveys}, specific knowledge about AI-supported QDA practices in these contexts remains scarce.

Another tension arises from the very foundations of AI itself: because many AI tools are grounded in statistical and probabilistic methods, their use in QDA may introduce assumptions that sit uneasily with interpretive qualitative inquiry~\cite{chatzichristos2025qualitative, paulus2024minutes, jiangSupportingSerendipityOpportunities2021}. This is amplified by the problem of overreliance on AI, facilitated by the plausibility of AI generations: in QDA, AI-generated "interpretations" or suggestions may be simply accepted, instead of being critically interrogated~\cite{kempny2025potenziale, okolo2022making, Buccinca2021}, creating lopsided structures of collaboration between AI and human researchers in QDA. This poses specific challenges for teams involving collaborators with varying degrees of literacy regarding AI as well as QDA, especially in an interdisciplinary field like HCI.
Thus, while AI holds potential to bridge ``big'' and ``small'' data approaches and to enhance efficiency, it raises equally urgent questions about reliability, ethics, and researcher agency. These risks and tensions require careful considerations for designing Human-AI collaboration in QDA, preserving the depth and richness of QDA while leveraging the potentials of AI.

\subsection{Designing for Human-AI Collaboration in QDA}
Integrating AI into \ac{QDA} is not simply a matter of automation, but of collaboration as well:
It requires careful design for human–AI and human-human collaboration in interpretive, context-sensitive work~\cite{jiangSupportingSerendipityOpportunities2021}.
Existing tools vary widely in how they support AI involvement, from minimal assistance to active roles in coding and analysis, with different levels of human involvement. 
Moreover, designing for researchers differing in proficiency in QDA, tool usage, as well as AI capability~\cite{lubars2019ask, feustonPuttingToolsTheir2021} poses specific challenges. 
For example, \citet{gauthier2022computational} developed a toolkit to facilitate computational thematic analysis, emphasizing the need to design it accessibly for a community with different proficiencies in tool usage as well as AI. Thus, they call for ethical designs to involve friction to remind researchers of the imperfection of computational models and issues concerning ethics and transparency. 
These challenges are also identified by \citet{kapania2025categorizing} who found HCI researchers struggling to identify such issues in their projects and needing guidance. 
Similarly, \citet{lam2024concept} suggest tools to be transparent to the researcher and acknowledge their limitations, \eg, by showcasing explicit inclusion criteria and scores.
At the same time, designs should avoid creating excessive ``unanticipated labor'' for researchers~\cite{feustonPuttingToolsTheir2021}.

While prior studies provide valuable insights into the potentials of AI in \ac{QDA}~\cite{taiExaminationUseLarge2024,morganExploringUseArtificial2023, rasheedCanLargeLanguage2024, lam2024concept, gauthier2022computational}, 
there is still limited understanding of how AI can be \textit{responsibly} integrated into researchers' real-life \ac{QDA} workflows, respecting their concerns and preferences~\cite{lubars2019ask, feustonPuttingToolsTheir2021}, facilitating collaboration between researchers as well as researchers and AI, while consolidating different research epistemologies. We are thus in need of comprehensive frameworks addressing these issues~\cite{owoahene2024review, kapania2025categorizing}. Moreover, it remains crucial to critically reflect not only on AI capability, but also what it \textit{should} do in QDA, preserving researcher concerns and preferences~\cite{feustonPuttingToolsTheir2021, lubars2019ask}. 
We extend this line of inquiry to the era of \acp{LLM}, identifying new steps, tasks, and different levels of involvement of AI to support QDA while maintaining methodological integrity and the human-centered values central to \ac{HCI} research practice.
To address these gaps, we study HCI researchers' \ac{QDA} workflows and their perspectives on the involvement of AI in qualitative research.
Specifically, we investigate the following research questions:

\begin{itemize}
    \item[\textbf{RQ1}] What do \ac{HCI} researchers' real-life \ac{QDA} workflows look like?
    \item[\textbf{RQ2}] What are \ac{HCI} researchers' concerns and conditions for integrating AI-based support into their \ac{QDA} workflow?
\end{itemize}

Our overarching goal is to \textit{explore how AI can be responsibly integrated into QDA, respecting the workflows, concerns, and conditions of HCI researchers.}
By systematically mapping QDA practices (RQ1) and surfacing researchers' conditions for AI integration (RQ2), we contribute a framework for AI support that is aligned with the real-world QDA workflows of HCI researchers. 
This framework synthesizes fragmented debates around QDA and maps the design space for AI tools, offering a structured vocabulary for researchers to reflect on if, where, and how AI may enhance -- rather than disrupt -- the collaborative and interpretive nature of qualitative research.

\section{Research Method}
\label{qdallm::sec::methods}
We conducted a two-step study to explore HCI researchers' perspectives on integrating AI into \ac{QDA}.
The first step involved semi-structured interviews with 15 \ac{HCI} researchers, representing a range of \ac{QDA} experience levels. 
These interviews were designed to uncover pain points, potentials, and conditions for AI support in \ac{QDA}, focusing on real-life workflows, challenges, and concerns.
To promote transparency and rigor, we preregistered our study on the Open Science Framework~(OSF) platform.\footnote{see \url{https://osf.io/aypgr/overview?view_only=1a27fda42755401981237e517838804a}} 
In the second step, we used the findings of the interviews to conceptualize ways \acs{AI} can be integrated into \ac{QDA}. 
We developed a framework to illustrate concrete examples of how AI can be integrated into the \ac{QDA} workflow and contextualize these examples with how they align (or do not align) with participants' concerns. 

{\small
\begin{table*}[t]
    \centering
    \caption{Detailed information on our participants' research backgrounds, their \ac{QDA} experience, and their primary \ac{QDA} tools and methods.
    Field categorizations (HCI, USP, Security) are based on participants’ self-description.}
    \label{tab:participants}

    \setlength{\tabcolsep}{2.5pt}
    \renewcommand{\arraystretch}{1.08}

    \begin{tabular}{@{}p{0.55cm}p{1.9cm}p{1.45cm}p{1.05cm}p{0.65cm}p{0.9cm}p{5cm}p{4cm}@{}}
        \toprule
        \textbf{No.} & 
        \textbf{Research Background} & 
        \textbf{Research Field} & 
        \textbf{Seniority} & 
        \multicolumn{2}{c}{\textbf{QDA Experience}} & 
        \textbf{Primary QDA Tools} & 
        \textbf{QDA Methods} \\
        \midrule
        P1 & CS / Sec. / Eng. & HCI, USP & ECR & 6 y. & 1--3 pr. & MAXQDA, Excel, Word, Google Docs/Sheets & TA, GT \\
        P2 & CS / Sec. / Eng. & HCI, USP & ECR & 3 y. & 4--6 pr. & MAXQDA & TA, CA \\
        P3 & CS / Sec. / Eng. & USP & ECR & 4 y. & 7--9 pr. & MAXQDA, NVivo, ATLAS.ti, Excel & TA \\
        P4 & CS / Sec. / Eng. & Security & ECR & 4 y. & 1--3 pr. & MAXQDA, Excel, Python & GT, Inductive Coding \\
        P5 & Social Science & HCI & ECR & 2 y. & 1--3 pr. & NVivo & Different kinds of TA \\
        P6 & Social Science & HCI & ECR & 3 y. & 1--3 pr. & MAXQDA, Excel, Word, Google Docs/Sheets & GT, Descriptive Analysis \\
        P7 & Social Science & HCI & ECR & 1 y. & 1--3 pr. & MAXQDA & CA \\
        P8 & CS / Sec. / Eng. & HCI & ECR & 1 y. & 1--3 pr. & MAXQDA & TA \\
        P9 & Social Science & USP & ECR & 3 y. & 4--6 pr. & MAXQDA, Excel & TA, CA \\
        P10 & CS / Sec. / Eng. & USP & ECR & 2 y. & 4--6 pr. & MAXQDA & TA, GT \\
        P11 & CS / Sec. / Eng. & HCI, USP & Senior & 6 y. & 7--9 pr. & NVivo, Excel, Google Docs & TA, Open Coding \\
        P12 & CS / Sec. / Eng. & USP & Senior & 9 y. & 10+ pr. & MAXQDA & TA, GT, Inductive Coding \\
        P13 & CS / Sec. / Eng. & Security & ECR & 1 y. & 1--3 pr. & MAXQDA, Excel & CA \\
        P14 & CS / Sec. / Eng. & HCI, USP & Senior & 3 y. & 4--6 pr. & Google Docs/Sheets, Python Scripts & TA, GT, Open Coding \\
        P15 & CS / Sec. / Eng. & USP & ECR & 3 y. & 1--3 pr. & MAXQDA, ATLAS.ti & Inductive Coding, Open Coding, TA \\
        \bottomrule
        \multicolumn{8}{@{}p{0.98\textwidth}@{}}{\footnotesize{\textbf{Legend.} CS = Computer Science, Eng. = Engineering, HCI = Human-Computer Interaction, Sec. = Security, USP = Usable Security and Privacy, ECR = Early Career Researcher, TA = Thematic Analysis, GT = Grounded Theory, y. = year(s), pr. = project(s).}}
    \end{tabular}
\end{table*}
}

\subsection{Study Materials}
Our study included a pre-questionnaire and a semi-structured interview with a sketching task.
The full interview guide is available in \autoref{qdallm::sec::appendix0}.
Both the complete pre-questionnaire and the interview guide are available in our OSF repository.

\xhdr{Pre-Questionnaire}
The pre-questionnaire consisted of eight questions, capturing participants' backgrounds and experience with \ac{QDA}. 
It covered their research background, current field of research, current position, tools used for \ac{QDA}, years of experience working with qualitative data, the number of research projects involving \ac{QDA}, and the data collection and analysis methods they have employed in the past.

\xhdr{Interview Guide}
The interview guide was organized into three parts:
\begin{inumerate}
    \item participants' real-life \ac{QDA} workflows
    \item pain points and gain points in \ac{QDA}
    \item perspectives on the potential for AI support\end{inumerate}

First (a), we explored the participants' overall research process, from data collection to analysis, with a particular focus on coding. %
We asked them to reflect on their most recent research project, discussing the types of data used, the methods and tools employed, resources accessed, and, if applicable, collaboration with other researchers.
To facilitate this reflection, we asked them to sketch their coding process on a piece of paper (\textit{sketching task}).
This visual representation not only helped them articulate their workflow but also served as a reference in subsequent explanations, allowing them to highlight specific areas where pain points arose or where they saw potential for AI support.
Then (b), we delved more deeply into the \ac{QDA} workflow, focusing on what participants found enjoyable or engaging, as well as what they found tedious or challenging.
We asked which steps were particularly time-consuming, the difficulties encountered, and the role of tool support, including both its benefits and limitations.

Lastly (c), we explored participants' expectations around AI-assisted \ac{QDA}.
We inquired about the conditions under which they would be willing to integrate AI into their workflows, which steps they believed could be automated, and which aspects should remain human-driven. 
Finally, participants were given the opportunity to share any additional thoughts on the topic.

\subsection{Recruitment \& Participants}
We used purposive and snowball sampling methods~\cite{palinkas2015purposeful} to recruit participants who had conducted at least one study involving \ac{QDA} through our professional networks within the HCI and Usable Security and Privacy (USP) communities.
We approached researchers directly and invited them to refer interested colleagues, leveraging peer identification to build a diverse and qualified sample.
Our recruitment strategy was designed to capture variation across research domains and methodological traditions relevant to HCI. 
We categorized participants’ primary research field based on their self-description.
To capture a broad range of experiences and perspectives, 
we took care to recruit participants from diverse group affiliations, different study approaches, topic choices, and educational backgrounds. 
Our goal was to achieve data saturation regarding participants' workflows, pain points, and gain points.

\xhdr{Sample Description}
\autoref{tab:participants} summarizes the participant sample.
We interviewed 15 HCI researchers with backgrounds in computer science, cybersecurity, engineering, and social sciences, all firmly embedded in HCI research, with some specifying to focus on USP.  
Twelve participants were early career researchers (mostly PhD candidates), and three were senior researchers. 
Eight had worked on 1-3 \ac{QDA} projects, four on 4-6, and three on more than~7 projects, with QDA experience ranging from 1 to 9~years. 
We recruited participants until data saturation was reached. We reached data saturation around interview 12 and conducted three additional interviews to confirm this.

Participants first completed the pre-questionnaire, which included an overview of the study procedure and a consent form. 
Afterward, we scheduled interviews. 
The interviews were semi-structured, conducted face-to-face, and lasted approximately 45 minutes.
Interviews were conducted in either English ($n=9$) or German ($n=6$), depending on participants’ native language and preference.
Two researchers were present: a lead interviewer who conducted the session and took notes, and a second researcher who created detailed protocols, including occasional verbatim quotes.
To protect participant privacy and follow data minimization principles, we chose not to record audio.
The lead interviewer, fluent in both German and English, conducted all interviews for consistency.
The second researcher varied among four researchers and was either fluent in English or fluent in both languages as needed.
To ensure completeness and consistency, they noted down points that stood out without performing any immediate analysis or interpretation. 
The researchers then compared, merged, and verified the completeness of their notes, ensuring no conflicts arose. 
Finally, when necessary, the researchers manually translated their synthesized interview protocols into English for subsequent data analysis.

We piloted our study with three participants to evaluate the procedure and interview flow. 
As no changes were made, these pilot sessions were included in the final dataset, resulting in 15 participants. 

\subsection{Ethical Considerations \& Data Protection}
The study was conducted in accordance with the Menlo Report's ethical principles~\cite{kenneally2012menlo} and approved by our institution’s ethics committee.
All participants gave informed consent via the pre-questionnaire and again at the start of the interview.
Before the interviews began, the interviewer walked the participants through the consent form and gave them the opportunity to ask questions. 
To protect the privacy of the participants, they were given pseudonyms after completing the questionnaire and all data was pseudonymized.
We stored and managed all data according to national data protection regulations and in accordance with our institution's policies on confidentiality and data protection.

\subsection{Data Analysis}
Three authors developed an initial set of deductive codes for tools and resources, data types, and participants' overall willingness to use AI, based on the pre-questionnaire response options and interview guide.
They each coded three protocols for the coding process, pain points, fun points, and concerns and conditions for using AI within \ac{QDA} to derive common themes.
This process involved closely examining the protocols to identify and conceptualize subcategories.
All three coders then created the final codebooks by discussing the identified themes and resolving conflicts in a collaborative session.
Two authors then coded the entire dataset, resolving disagreements through discussion until consensus was reached.
To help us later illustrate our findings, the coders included example quotations from the protocols in our codebooks, both paraphrased and verbatim.

Additionally, we analyzed the drawings from the sketching task to derive commonalities and differences in the coding process and extracted participants' suggestions for concrete AI applications as in vivo codes.
Two coders reviewed sketches together with protocols to create a combined metaview of current QDA practices. 
Each participant’s workflow was redrawn and sequentially mapped into a shared metaview.
New elements were added as they emerged, recurring ones were skipped.
Closely related processes were later merged to reduce complexity. 
We differentiated elements taking into account the SeeMe-method developed by Herrmann \cite{herrmann2012kreatives}.
SeeMe is a semi-structured socio-technical modeling technique used for task analysis that involves three visualization elements: (1)~roles assigned to individuals or teams, (2)~activities that are of a more dynamic characteristic and represent change functions, and (3)~entities, defined as objects from the physical world. For entities, we excluded tools and software support from the metaview as they were heavily dependent on the participant's individual preference, but generally involved in each process element.

\section{Results}
\label{qdallm::sec::results}
Here, we present HCI researchers' \ac{QDA} workflows, including their pain points (\autoref{qdallm:results:rq1}) and report participants' willingness to use AI support along with concerns and conditions (\autoref{qdallm:results:rq2}).
In \autoref{qdallm::sec::appendix1}, we provide the complete codebooks from our interviews, including descriptions, examples, and quantities for each code. \autoref{qdallm::sec::appendix2} shows example workflow sketches.

\subsection{RQ1: \acs{HCI} Researchers' Real-Life \acs{QDA} Workflow}
\label{qdallm:results:rq1}

\subsubsection{Contexts of Studied \acs{QDA} Workflows}

In the pre-questionnaire, participants reported having analyzed open-ended survey responses ($n=12$), interview transcripts ($n=11$), written documents ($n=8$), and social media posts ($n=5$) as the most common data types. 
Most participants ($n=9$) reported using more than one tool for qualitative coding, including \textit{MAXQDA} ($n=12$), Excel ($n=6$), and Google Docs ($n=4$). 
Participants also indicated experience with multiple methodological frameworks, often adapting them based on the nature of the data and their research context.
\textit{Thematic Analysis} ($n=11$), \textit{Grounded Theory} ($n=6$), and \textit{Content Analysis} ($n=4$) were mentioned most frequently.

\subsubsection{Uncovering Participants' Actual Coding Workflow}

\begin{figure*}[t]
    \centering
    \includegraphics[width=\linewidth]{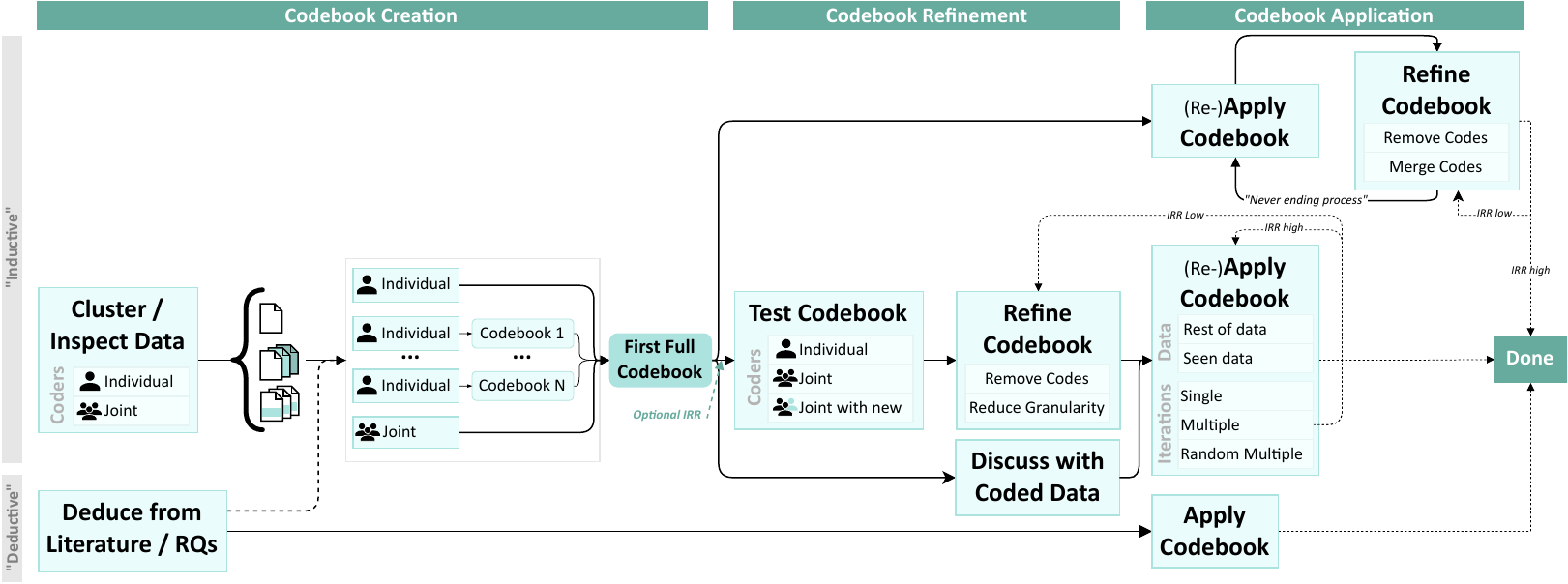}
    \caption{Detailed overview of participants' real-world coding workflow, consisting of three stages for codebook creation, codebook refinement, codebook application, and different steps taken in inductive and deductive approaches.}
    \Description{The figure shows a flowchart depicting the 3 stages of the coding workflow: Codebook Creation, Codebook Refinement, and Codebook Application. It has 2  starting points: 1 for inductive, and 2 for deductive, and 1 endpoint labeled Done. 12 elements depict different steps in this process, including visuals representing individual or joint coding. 2 types of lines connect the different elements: arrows depict the element as next step. Dotted arrows depict the next element as optional. The elements representing steps are listed below.
    Starting point 1: Inductive
    Step 1: Cluster/Inspect Data. Individual or joint. From this, an arrow leads through pictograms of different data sets to step 2.
    Step 2: An element with multiple pictograms labeld Individual and 1 labeled Joint are depicted. They are connected to elements labeled Codebook 1 to Codebook N. An arrow leads to the step 3.
    Step 3. An element labeled First Full Codebook is shown at the end of the stage Codebook Creation. An arrow leads to 3 different elements: Step 3.1, Step 3.2, and Step 3.3. A dotted arrow leads to this arrow, labeled Optional IRR.
    Step 3.1 (Re-)Apply Codebook in the Stage of Codebook Application. An arrow leads to the next element: Refine Codebook, including labels for Remove Codes and Merge Codes. A dotted arrow labeled IRR high leads to the final step: Done. From this dotted arrow, another arrow with the description IRR low leads back to the element Refine Codebook. Another arrow labeled never ending process leads back to the element (Re-)Apply Codebook. 
    Step 3.2: Test Codebook in the Stage of Codebook Refinement. This element consists of 3 pictograms for coders. 1: Individual, 2: Joint, 3: Joint with new. An arrow leads to step 4.
    Step 4: Refine Codebook in the Stage of Codebook Refinement. The element contains labels for Remove codes and Reduce granularity. An arrow leads to step 5.
    Step 5. (Re-)Apply Codebook in the stage of Codebook Application. It contains labels for Rest of data and Seen data. It further shows iterations labeled Single, Multiple, or Random Multiple. A dotted arrow labeled IRR high leads back to the same element (Re-)Apply Codebook. A dotted arrow labeled IRR low leads back to step 4: Refine Codebook. Another dotted arrow leads to the last element: Done.
    Step 3.3: An element labeled Discuss with Coded Data. From here, an arrow leads to step 5: (Re-)Apply Codebook.
    Starting point 2: Deductive
    Step 1: An element labeled Deduce from Literature / Research Questions. A dotted arrow leads to Step 2 in Inductive Coding. A straight arrow leads to the next step in Codebook Application.
    Step 2.2: An element labeled Apply Codebook. A dotted arrow leads to the last element: Done.}
    \label{fig:workflow}
\end{figure*}

At a high level, our participants' coding processes generally followed three main steps:
\begin{enumerate*}
    \item \textbf{Codebook Creation} -- developing an initial version of the codebook,
    \item \textbf{Codebook Refinement} -- updating the codebook through discussion, or by applying it iteratively to new, unseen data, and
    \item \textbf{Codebook Application} -- applying the codebook to the full dataset, with occasional ongoing adjustments.
\end{enumerate*}
As illustrated in \autoref{fig:workflow}, the majority of our participants followed an \textit{inductive} approach. 
Only two created their codebooks based on a priori information with \textit{deductive} coding. 
Team structures varied: some participants coded alone, while most had a lead researcher handle codebook \textit{creation} and \textit{refinement}, with team input during discussions. In some cases, new researchers applied the final codebook to the full dataset.

Typically, participants began by reading through a subset of the data to develop an overview.
Depending on the project, either the entire dataset or a specific subset (often around 10\% of the data) was then used to develop an initial codebook.
This was done either individually, in parallel by multiple coders who later merged their versions, or collaboratively from the start.
In certain instances, this codebook was then applied to the rest of the data, followed by iterative codebook refinements, referred to as a ``\textit{never-ending process},'' as the codebook application progressed and new themes emerged or began to overlap.
However, most reported iteratively applying the codebook to new small sets of data and refining the codebook along the way by removing codes and reducing granularity until all team members were satisfied with the result.
Finally, the codebook was applied to the rest of the data. This allowed the determination of a final code count.

There was an almost equal split between those who mentioned that the codebook was only applied by one researcher and those who conducted multiple rounds of coding to be able to calculate inter-rater reliability~(IRR).
In some cases, IRR was used during early codebook development to support refinement; in others, it was calculated only at the end for reporting purposes.
One participant described using random checks for verification, rather than full double-coding.

\subsubsection{Pains and Gains}
Researchers discussed several challenges and difficulties (``\textit{pain points}'') and rewarding aspects (``\textit{fun points}'') of their QDA workflows.
Understanding both helps identify opportunities for AI support that align with researchers' values and practices.

\xhdr{Pre-processing} %
Data often needs to be pre-processed, \eg, transcription of audio recordings. 
Some participants found this a boring and tedious task, especially when the transcriptions had to be manually corrected or segmented. 
Several participants expressed particularly enjoying the process of familiarizing themselves with the processed data and finding valuable insights and learning opportunities.
This deep engagement was often highlighted as a rewarding aspect of \ac{QDA}.

\xhdr{Codebook Creation and Refinement}
Developing the codebook was often described as time-consuming and presented several challenges:
\begin{enumerate}
    \item \textbf{Deciding Focus \& Initial Exploration.}
    Deciding the focus of the codes was a key difficulty, as they need to align with the RQs and the specific project context. 
    A few participants noted that poorly defined RQs can lead to incoherent codes. 
    The initial stages of coding were described as particularly challenging: ``\textit{Going through the first data segments can go wrong in so many ways, and then you have to do it all over again. You learn, but you don’t get a result from that.}'' (P6)
    \item \textbf{Naming and Refining Codes.}
    Naming and refining codes also requires substantial effort, as codes must be concise and distinguishable: ``\textit{It can be a complicated process, trying to encapsulate something into a code that also applies to other samples. Going back and refining codes can be very iterative, annoying, and a big challenge.}'' (P3). This challenge was particularly prevalent among less experienced researchers.
    \item \textbf{Managing the Codebook.}
    Several participants reported difficulties in managing the codebook's complexity as it grows, having concerns about losing sight of the big picture and ensuring that codes remain coherent.
    Inferring hierarchies also posed challenges because 
it can be ``\textit{painful to put them into high-level themes without missing an important point.}'' (P1). This pain point was particularly pronounced among participants who handled large datasets or repetitive coding tasks.
    \item \textbf{Restructuring the Codebook.}
    Most participants expressed concerns about restructuring the codebook. 
    They dreaded having to go back and adapt codes, update the codebook, or even start anew if the codebook ended up not fitting the data well: 
    ``\textit{where the codebook begins is very different from where it ends.}'' (P5)
\end{enumerate}
Despite these struggles, several participants acknowledged that the process of developing a codebook is highly educational. 
Synthesizing patterns and refining themes was also described as exciting and rewarding, offering a sense of accomplishment.

\xhdr{Codebook Application} 
Participants described codebook application as one of the most time-consuming tasks.
They expressed frustration when adjustments to the codebook were necessary later in the process: ``\textit{When new codes were identified, the codebook had to be updated, and already-coded data had to be revisited to determine if the new code applied.}'' (P15)
Nonetheless, nearly half of the participants perceived the process of applying the codebook to the data as satisfying. 

\xhdr{Tools}
Throughout the \ac{QDA} process, participants use a variety of tools, some of which they report having difficulty with. 
For one participant, managing multiple tools was burdensome, as each tool specialized in different tasks, requiring regular data transfers between them.
Issues included cumbersome data transfers, file management, import/export limitations, and collaboration features. In particular, \textit{MAXQDA}’s complexity overwhelmed some users.

\xhdr{Lack of Structure and Guidance} Participants with limited \ac{QDA} experience often expressed concerns about the lack of structure and guidance in \ac{QDA}. 
They reported feeling uncertain about whether they were conducting the analysis correctly, expressing a need for assistance, such as having a second person review their work.
This was linked to the perception that coding is subjective and inherently messy, making it difficult to evaluate whether their observations had objective value or were merely personal insights. Here, collaboration with other, more experienced researchers was brought up as helpful.

\xhdr{Post-Processing} 
Participants valued the final stages of analysis, particularly the synthesis of patterns and the formulation of initial results, as one of the most fulfilling parts of the process. 
This stage allowed them to see the culmination of their efforts and begin to draw conclusions from the data.
Additionally, some participants said they enjoyed writing the report and communicating their findings, as well as
extracting interesting quotes for the manuscript.

\xhdr{Collaboration Across Stages} 
Most participants reported that while collaboration and iterative discussions are essential to their \ac{QDA} process, those can be tedious and frustrating, particularly in cases where multiple interpretations are possible or different backgrounds are present, making it difficult to reach consensus.
Participants emphasized that different analysts may apply categories based on different underlying assumptions (\textit{“understand categories differently,”} P15), and that negotiating these differences can make the codebook \textit{“explode”} rather than converge on shared definitions (P7). 
In several workflows, this alignment work involved iterative loops of independent coding followed by comparison meetings and codebook revisions; in some cases, this was driven by expectations around IRR, described as requiring very frequent coordination (e.g., meeting \textit{“nearly every day to get some IRR,”} P2). 
Where IRR was calculated, participants reported that it could be discouraging; ranging from \textit{“only 50\%”} agreement that revealed diverging interpretations of the same code (P8) to experiences of extremely low agreement (\textit{“cruel 20--30\%,”} P7). 

A few participants described onboarding others and initial discussions as cumbersome.
This included the effort of briefing and calibrating how codes should be applied (P2) as well as the broader challenge that new coders may lack full project context (P1). 
Some teams addressed disagreement by involving a third person as a tie-breaker (P10; P12).
More generally, dependency risks were also reported: relying on another person can mean delays if they drop out or do not deliver the expected quality (P12). 

Several accounts also stressed that tooling can amplify friction: collaboration was described as difficult when researchers cannot work synchronously in the same project (P1; P7) or when sharing/merging is unstable or \textit{“a disaster”} (P5), leading teams to rely on workarounds such as Google Sheets/Word memos to coordinate and document decisions (P1; P11; P4).

On the positive side, participants highlighted the value of sharing their thoughts with others and thereby learning to better understand the data and the QDA process. 
These discussions were seen as enriching and helping to refine ideas.
Some explicitly described the interpretive exchange as enjoyable (\textit{“the discussion makes [it] fun,”} P15; \textit{“I enjoy the process of discussing because that gives you another perspective,”} P6), and valued collaboration as a second pair of eyes to improve clarity and consistency (P1; P9).

\begin{summarybox}
\small
\subsubsection*{Answering RQ1: Real-life \ac{QDA} workflows exhibit structural similarities but vary with project specifics}
Our study provides a detailed view of how \ac{HCI} researchers perform and perceive \ac{QDA} in their everyday practice.
Our interviews revealed that while participants' workflows are similar in structure, they vary highly depending on the project, methodology, data type, approaches in which codebooks are created, and degrees and quality of collaboration.
As \autoref{fig:workflow} illustrates, deductive coding is a somewhat straightforward process, whereas inductive coding, more data-driven and interpretative, involves several loops. 
This is also mirrored in the pain points participants highlighted and has implications for the integration of AI into the \ac{QDA} workflow, as highly context-specific use cases also need to be addressed. 
Similarly, other perceived ``pains and gains'' during the workflow are often not mutually exclusive: what one researcher considers as enjoyable and vital to the research can be another researcher's biggest annoyance. For example, some participants described iterative discussions to reach consensus as tedious, while others experienced them as educational.
Less experienced researchers expressed a need for guidance and structure in their workflows, reflecting the importance of mentorship and learning in QDA facilitated by collaboration with others.
This highlights the importance of building AI assistants in a configurable and customizable manner, not as a one-size-fits-all solution, but as systems that can adapt to researchers' varying needs and levels of expertise.
\end{summarybox}

\subsection{RQ2: Participants' Concerns and Conditions for Integrating AI Support}
\label{qdallm:results:rq2}
Since participants often used \textit{``AI''} broadly when discussing QDA tools, we preserve this broader term when reporting their perspectives and use \textit{``LLMs''} when the underlying technology or cited work is specifically language-model based.

Almost all participants were open to exploring AI tools in their \ac{QDA} process.
However, most of these participants emphasized that certain conditions should be met or that they would still want to have full control over the \ac{QDA} process. 
While a few participants were eager to integrate AI into their workflow, one was strongly opposed to the use of AI, stating that they did not see any useful applications for it in their workflow. %

Several participants recognized the potential of AI to automate repetitive tasks -- such as data pre-processing, organizing raw data, or annotating thousands of survey responses.
Others saw an opportunity for AI to contribute ideas and enrich the analytical process with new views and perspectives. 
They expressed interest in AI's ability to identify relationships and suggest novel patterns in the data that might be overlooked, complementing human analysis.
However, they also expressed various concerns, which we categorized into four main themes: \textit{research ethics}, \textit{personal involvement and responsibility}, \textit{quality of AI output}, and \textit{specific requirements of AI tools}.
Across our interviews, we did not observe a clear pattern linking participants' level of experience in \ac{QDA} to their willingness to adopt AI or their concerns regarding its integration. 
Both novice and experienced researchers expressed openness to exploring AI tools while emphasizing the importance of safeguards and customization to align with their workflows.

\subsubsection{Research Ethics}
Many participants expressed concerns about the ethical implications and scientific rigor associated with introducing AI into the research process. 
Specifically, they stressed concerns about privacy, transparency, and confidentiality: participants emphasized the need for AI systems to comply with data protection regulations such as the GDPR, ensuring data storage and usage are transparent and secure, especially when third-party model providers were involved. They wanted clear information on where the data would be stored and how it would be used by the companies providing the AI services. 
One participant suggested research participants should be informed before data collection if AI will be used to analyze their data. 
Another participant insisted that any data fed into an AI system must be strictly anonymized, with all personal identifiers removed to protect participant privacy. 
One participant expressed fear that integrating AI into their research could potentially compromise the confidentiality of their work, risking that AI would expose novel insights or ideas derived from the data.
Further, participants were worried about the handling and destination of the data, especially regarding \textit{sensitive data and topics}, noting that AI systems require special supervision when dealing with sensitive topics, as the involvement of AI might shape the interpretation and dissemination of knowledge in these domains.

\subsubsection{Personal Involvement and Responsibility}
Participants were reluctant to become overly dependent on AI tools, emphasizing the need to remain directly connected with their data and engaged with the analysis. 
Even with AI assistance, researchers must retain responsibility for their research outcomes. 
This sentiment reflects a concern for preserving the integrity and independence of academic research, ensuring that AI serves as a supportive tool rather than replacing the researcher’s role.
Moreover, conducting \ac{QDA} involves learning from the process and improving through experience. 
As one researcher explained, ``\textit{doing the coding has helped me immensely with the analysis, I think it is necessary and I would not want to give that up}''.
Participants agreed that AI tools should enhance human capabilities rather than degrade them or make the process more cumbersome.
One participant expressed concerns about job displacement, cautioning against AI fully replacing human coders.

\subsubsection{Quality of AI Output}
Participants expressed uncertainty about the capabilities of AI tools, particularly regarding the reliability and trustworthiness of their outputs. 
A key concern mentioned by several participants was the potential for AI to produce false information, which could compromise the integrity of research.
They emphasized that the consequences of such errors could propagate into the final analysis and published findings. %
Participants were also wary of an AI system's ability to handle qualitative data, noting that it lacked the necessary experience and background for such tasks: 
``\textit{I see problems with validity when working with AI. [...] The researcher has a learning process and will apply categories differently over time, AI does not.}''
This skepticism led to a reluctance to fully rely on AI for data analysis: 
``\textit{I can't imagine that there is such an [AI] algorithm and that a human doesn't have to check it.}''
Concerns about biases that AI might introduce further fueled this mistrust: 
``\textit{If we add bias to it, we can as well use mine; at least then I know what went wrong, and can account for it in the limitations.}''

\subsubsection{Conditions and Requirements}
Given their concerns, participants formulated conditions and requirements for the potential use of AI in \ac{QDA}, emphasizing tools would need to be transparent, customizable, and supportive of human oversight.
To meet privacy considerations, a few participants stressed that AI models should be able to operate in offline mode, ensuring that no data would be uploaded or shared with third parties. 
Addressing the need for personal involvement and responsibility, several participants emphasized the importance of evaluating the performance and reliability of AI tools before fully integrating them into their workflows. 
Many participants were particularly concerned about retaining control and agency over the analysis process -- with several participants stressing the ability to adjust the level of AI’s influence, preferring a more interactive collaboration with AI rather than a fully automated analysis. 
A majority of participants also expressed a strong desire for mechanisms to check and verify AI outputs, ensuring the results could be trusted.
Several participants further wanted explanations for AI-generated codes: 
``\textit{I would want an explanation for each code, or at least an example from the data to show where the decision was coming from.}'' 
Other participants wanted to take random samples of AI outputs and check them or compare them to their own results. 
Generally, many participants expressed a preference for an interactive approach to AI integration, where they could collaborate with the AI rather than relying on it entirely. 
Some participants welcomed the possibility to modify AI settings or tune outputs to their needs.

\begin{summarybox}
\small
\subsubsection*{Answering RQ2: \ac{HCI} researchers are open to AI support if it enhances their workflow while addressing ethical, autonomy, and quality concerns}

Participants were generally open to AI in \ac{QDA}, provided it enhanced rather than replaced their skills.
They valued AI for offloading repetitive tasks, enabling them to focus on deeper interpretive work. 
Others saw an opportunity in AI to enrich the analysis with new views and perspectives.
Hence, AI was perceived both as a tool for efficiency and as a catalyst for analysis.
Participants expressed several concerns 
and mentioned conditions that had to be met.
They suggested developing AI tools that operate offline to allow for greater control over the data's use and storage.
Quality and trustworthiness of AI outputs were another major concern, especially given AI's current limitations in dealing with nuanced data and the risk of generating false information.
Participants stressed the importance of rigorously testing and validating AI tools before integrating them into their research workflows to ensure their reliability in research contexts.
Still, they saw potential in AI tools that are developed in alignment with their needs, with a focus on transparency, control, and ethical compliance. 
Generally, they preferred AI to support collaborative sensemaking, whether in engagement with the data or other researchers, rather than replacing it.
Notably, willingness to use AI did not appear to track with participants’ experience level, suggesting these conditions are shared across different levels of expertise.
\end{summarybox}

\section{A Framework for \acs{AI}-based Support that Aligns with \acs{HCI} Researchers' Real-Life \acs{QDA} Workflows}

\label{qdallm:results:rq3}

Our results indicate a general openness to exploring the potential of AI support for QDA  among HCI researchers.
However, this willingness is balanced by significant concerns about research ethics, the quality of AI output, and maintaining control over the research process.
Participants have expressed different suggestions and ideas for incorporating AI in their workflows.
Our interviews revealed that researchers do not conceptualize AI support in \ac{QDA} as a binary choice between automation and human interpretation.
Instead, participants articulated a spectrum of acceptable AI involvement, ranging from complete exclusion to conditional or near-autonomous delegation.
Based on these insights, we propose a tiered framework that characterizes AI integration along increasing levels of agency, epistemic influence, and autonomy.
To develop the framework, we extracted participants' concrete suggestions for AI integration as in vivo codes and grouped them by the degree of AI agency they implied -- from tasks where AI has no influence on analytic outcomes to tasks where AI acts with varying degrees of autonomy. 
We chose AI agency as the organizing dimension because participants themselves consistently framed their openness and concerns in terms of how much control AI would have over their analysis. 
We then situated each role within related work and existing tool implementations to provide tangible design examples. 
We acknowledge that some roles could reasonably span multiple tiers depending on implementation; the assignment reflects the typical level of AI agency implied by participants' descriptions. 
Importantly, the tiers describe increasing potential AI agency but are not intended as a progression that researchers should follow sequentially. 
Researchers may draw from multiple tiers simultaneously within a single project -- for instance, using AI autonomously for transcription while also engaging it as a sparring partner during codebook refinement. 
Because participants primarily described desired capabilities rather than specific model architectures, we use AI as an umbrella term in the framework. The roles therefore differ in their technological scope: some can be implemented with non-LLM methods, such as transcription, IRR calculation, topic modeling, or frequency-based analytics, while others are more directly LLM-enabled, such as conversational sparring, explanation, codebook generation, or open-ended coding suggestions.
We group the identified roles into three levels of AI involvement -- minimal, moderate, and high (see \autoref{tab:ai-qda-framework}).

\begin{table*}[ht]
\small
\centering
\caption{A tiered framework for integrating AI into \acf{QDA}, based on participant interviews, with three stages ranging from minimal to high AI involvement.}
\label{tab:ai-qda-framework}

\setlength{\tabcolsep}{3pt}
\renewcommand{\arraystretch}{1.15}

\begin{tabular}{@{}p{1.35cm}p{4cm}p{2.5cm}p{2cm}p{6.5cm}@{}}
\toprule
\textbf{Tier} & \textbf{AI Role} & \textbf{Origin} & \textbf{Related Work} & \textbf{Participant Quote} \\
\midrule

\textbf{Minimal}
& Human-only 
& P3 
& \cite{davisonEthicsUsingGenerative2024}
& ``Naturally, I want to do most of the things myself'' \\

& AI for basic technical tasks 
& P2, P4, P6, P7, P12 
& \cite{whisperaiWhisperAIa} 
& ``I am very happy about automated transcriptions'' \\

& AI as an intelligent user interface 
& P1 
& \cite{OfficeAssistant2025} 
& ``Some tools are really difficult to get to know.'' \\

& AI as a productivity tool 
& P2 
& \cite{SecureScalableAI,RayyanAIPoweredSystematic} 
& ``Calculating IRR in MAXQDA can be a big pain.'' \\

& AI as a training tool 
& P4, P1 
& \cite{paleaAnnotaPeerbasedAI2024,DedooseMakesQualitative,wang2019human} 
& ``Labeling exercises'' \\

\midrule

\textbf{Moderate}
& AI as a mediator 
& P4 
& \cite{drouhardAeoniumVisualAnalytics2017,gao2024collabcoder,ganjiEaseCodeComplex2018}
& ``Comparison of different codebooks'' \\

& AI for validation 
& P12, P4 
& \cite{morganExploringUseArtificial2023}
& ``You could use it to validate your own ideas---and still discard it.'' \\

& AI for refinement and explanations 
& P1, P2, P8 
& \cite{MAXQDAOfficialSite} 
& ``AI can help to get some structure into my codebook'' \\

& AI for analytics 
& P12, P5, P10, P9, P4 
& \cite{paleaAnnotaPeerbasedAI2024}
& ``Give me warnings if I am too detail-oriented.'' \\

& AI suggestions 
& P12, P8, P11, P4 
& \cite{maratheSemiAutomatedCodingQualitative2018,rietzCodyAIBasedSystem2021,gao2024collabcoder}
& ``The tool suggests an important code or a frequent word.'' \\

& AI as a sparring partner 
& P7, P10, P11, P9 
& \cite{chew2023llm}
& ``As a sparring partner, I can imagine this in all steps'' \\

\midrule

\textbf{High}
& AI with human-in-the-loop 
& P8, P4 
& \cite{cohnHumanLoopLLMApproach2024,daiLLMloopLeveragingLarge2023a}
& ``The AI suggests a codebook, and I give feedback'' \\

& AI with human approval 
& P13, P12, P8 
& \cite{drapalUsingLargeLanguage2023,xiaoSupportingQualitativeAnalysis2023,banoAIHumanReasoning2023}
& ``AI could apply my codebook, and I review a random subsample'' \\

& AI with conditional autonomy 
& P4 
& \cite{lennonDevelopingTestingAutomated2021}
& ``I see AI as tooling for low-hanging fruit, so I can focus on more complex codes'' \\

& Full AI delegation (partial or full) 
& P1, P8 
& \cite{lam2024concept,rasheedCanLargeLanguage2024,AccuratelyAnalyzingLarge2024}
& ``The tool could automatically create a codebook.'' \\

\bottomrule
\end{tabular}

\end{table*}

\subsection{\textbf{Minimal \acs{AI} Involvement}: Supporting Human-Centric Workflows}
\label{subsec:minimal}
At this level, AI has little to no direct impact on the core \ac{QDA} workflow and primarily serves as a supportive tool.
We emphasize this tier as an important baseline rather than a limitation, as it highlights that AI-supported QDA must remain optional and that full human control continues to be a legitimate and valued practice.

\xhdr{Human-Only}
At the lowest level, the researcher conducts the entire \ac{QDA} process without AI assistance.
This approach is particularly relevant when working with sensitive data, highly contextual material, or when researchers lack trust in AI’s ability to achieve nuanced interpretation or simply do not have access to suitable tools.
While one participant was reluctant to use AI altogether (P3), others were open to involving AI.
Given longstanding concerns about opacity, ethical risks, and the erosion of interpretive depth in AI-supported qualitative work~\cite{khalili2023against,ashwin2023using,hitch2024artificial}, this level remains crucial.

\xhdr{AI for Basic Technical Tasks}
More commonly, participants accepted AI for repetitive, non-interpretive tasks such as transcription, formatting, or data cleaning.
As P12 noted, ``transcription can be done by AI as long as it complies with our privacy practices''.
For instance, tools like Whisper~\cite{whisperaiWhisperAIa} can transcribe audio recordings.

\xhdr{AI as an Intelligent User Interface}
AI may also act as a usability layer within QDA software, guiding researchers through complex features without shaping analytic outcomes.
P1 explained, ``some tools are difficult to get to know'' and likened this role to metaphors such as Microsoft's Clippy \cite{OfficeAssistant2025}, providing domain knowledge and maintaining a model of the user.

\begin{figure}[!htbp]
\centering
\includegraphics[width=.99\linewidth]{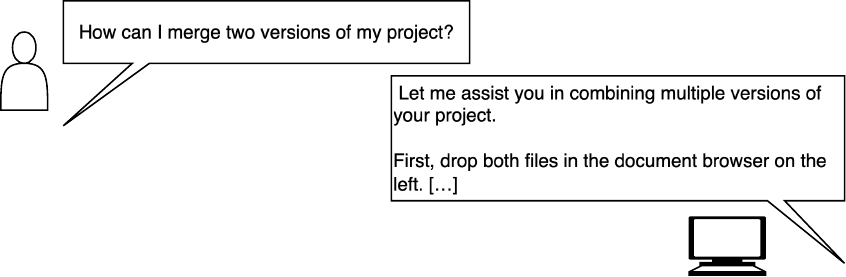}
\par
\caption{[AI as an Intelligent User Interface] \textnormal{AI provides step-by-step guidance within QDA software, suggesting functionalities based on user behavior and answering questions to support with problems in tool usage.}}
\Description{Figure 2 visualizes the role AI as an Intelligent User Interface. A pictogram of a person with a speech bubble asks: How can I merge two versions of my project? A pictogram of a computer with a speech bubble answers: Let me assist you in combining multiple versions of your project. First, drop both files in the document browser on the left. [...].}
\label{fig:ai_iui}
\end{figure}

\xhdr{AI as a Productivity Tool}
In this role, AI enhances productivity by automating evaluative tasks, such as summarization, translation, literature screening, or inter-rater reliability (IRR) calculations.
P2 highlighted that \textit{``calculating IRR in MAXQDA can be a big pain. AI could offer a customizable IRR calculation''}, suggesting AI could reduce overhead.
Tools like DeepL\cite{SecureScalableAI} and Rayyan\cite{RayyanAIPoweredSystematic} illustrate how AI already demonstrates its utility beyond QDA.

\begin{figure}[!htbp]
\centering
\includegraphics[width=.8\linewidth]{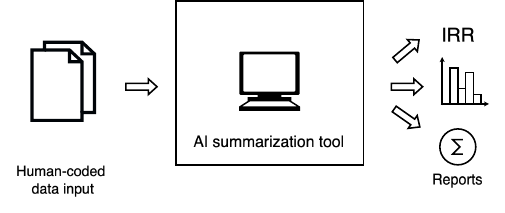}
\par
\caption{[AI as a Productivity Tool] \textnormal{AI generates summary reports of QDA projects, such as identifying the most or least frequent codes, or calculating IRR, allowing the researcher to review these insights and make informed decisions more efficiently.}}
\Description{Figure 3 visualizes the role AI as a Productivity Tool. A pictogram shows 2 paper sheets, labeled  Human-coded data input. An arrow leads to a symbol of a computer, labeled AI summarization tool. From here, 3 arrows lead further. Arrow 1 leads to IRR. Arrow 2 leads to bar graph pictogram. Arrow 3 leads to a sum pictogram, labeled Reports.}
\label{fig:ai_productivity}
\end{figure}

\xhdr{AI as a Training Tool}
Finally, AI can support onboarding and training, particularly for less experienced qualitative researchers.
P4 mentioned the idea of \textit{``labeling exercises''}, guided practice to build confidence and consistency.
For example, the Dedoose\footnote{https://www.dedoose.com/userguide/interraterreliability} training center supports research teams in maintaining coding consistency. 
Palea et al.~\cite{paleaAnnotaPeerbasedAI2024} propose a tool that generates AI hints to help students improve their coding by understanding annotation errors. Wang et al. show AI can be used as a teaching tool in data science~\cite{wang2019human}.

\begin{figure}[!htbp]
\centering
\includegraphics[width=.8\linewidth]{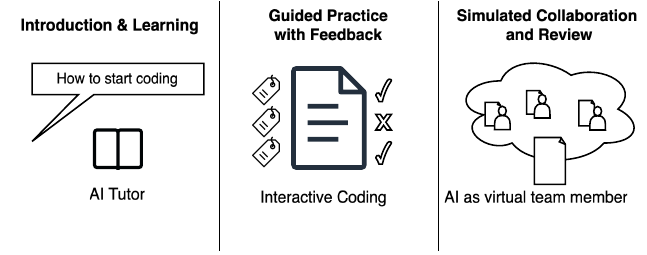}
\par
\caption{[AI as a Training Tool] \textnormal{AI supports new researchers in learning and practicing QDA through personalized, interactive instructions, real-time feedback, and simulated collaboration exercises to develop critical thinking and improve coding consistency.}}
\Description{Figure 4 visualizes the role AI as Training Tools. 3 visualizations are separated by lines, depicting different examples. 
The first is titled Introduction and Learning. A speech bubble asks: How to start coding? A pictogram of a book is labeled as AI Tutor. 
The second is titled Guided Practice with Feedback. A pictogram of a paper sheet is framed by 3 pictograms of tags on the left, and 2 checkmarks and 1 X on the right. It is labeled Interactive Coding.
The third is titled Simulated Collaboration and Review. A cloud-shaped bubble contains 3 pictograms of persons with paper sheets. A big paper sheet pictogram is depicted at the bottom. It is labeled AI as virtual team member.}
\label{fig:ai_training}
\end{figure}

\subsection{\textbf{Moderate AI Involvement}: Human-AI Collaboration}
\label{subsec:moderate}
At this level, AI works alongside human researchers to support and complement their coding efforts.
The product of \ac{QDA} is still based on human research efforts and revolves around their expertise and interpretation.

\xhdr{AI as a Mediator}
Collaboration was envisioned by participants as being supported by AI by surfacing disagreements, aligning diverging codebooks, and focusing discussions during consensus finding. Rather than resolving meaning, AI highlights where interpretation diverges, reducing coordination costs.
For instance, Aeonium~\cite{drouhardAeoniumVisualAnalytics2017} highlights ambiguities and inconsistencies between multiple coders by providing visual overviews of assigned codes. %
CodeWizard~\cite{ganjiEaseCodeComplex2018} identifies coders' certainty and disagreements to improve consensus.
CollabCoder~\cite{gao2024collabcoder} provides real-time synchronization, acting as a suggestion provider, mediator, and facilitator throughout the workflow.
Participants also imagined AI support for collaboration as a \emph{glitch-free} shared workspace with reliable real-time synchronization, scaffolding for onboarding and early calibration of new team members, and as an additional collaborator or ``third instance'' to spar with (P4; P6). The latter included producing tentative codes to compare against human coding (P7). 
Still, they expressed skepticism that AI could replace theme-building collaborative conversations (P3) and questioned whose labels an AI should learn from when research teams already show low agreement (P12).

\begin{figure}[!htbp]
\centering
\includegraphics[width=.8\linewidth]{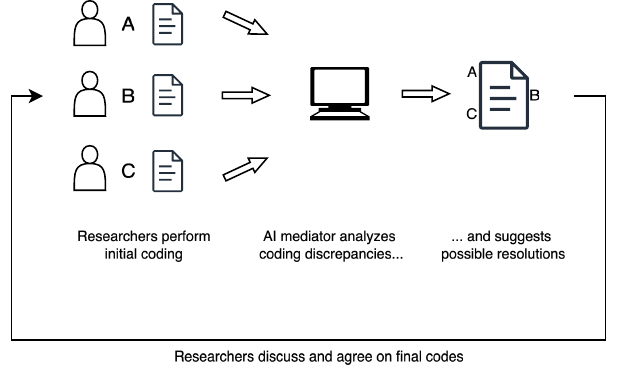}
\par
\caption{[AI as a Mediator] \textnormal{AI helps researchers resolve coding conflicts and reach consensus efficiently by analyzing discrepancies and suggesting resolutions.}}
\Description{Figure 5 visualizes the role AI as a Mediator. It depicts a small flowchart with 3 steps. Step 1 is labeled Researchers perform initial coding. It contains 3 pictograms of persons, labeled A, B, and C, each with a pictogram of a paper sheet. From each person, an arrow leads to step 2, a pictogram of a computer, labeled AI mediator analyzes coding discrepancies ... . An arrow leads to step 3: A pictogram of a paper sheet framed with letters A, B, and C. It is labeled  ... and suggests possible resolutions. From here, an arrow leads back to step 1. The arrow is labeled Researchers discuss and agree on final codes.}
\label{fig:ai_mediator}
\end{figure}

\xhdr{AI for Validation}
Participants, especially those less experienced, expressed uncertainty about their coding decisions and saw potential in using AI for validation, \eg, by comparing their results with AI-identified codes or segments.
In this role, AI serves as a validation tool, identifying gaps or inconsistencies.
For instance, AI could verify the completeness of coding and highlight passages that may need revisiting (P4).
\begin{figure}[!htbp]
\centering
\includegraphics[width=.9\linewidth]{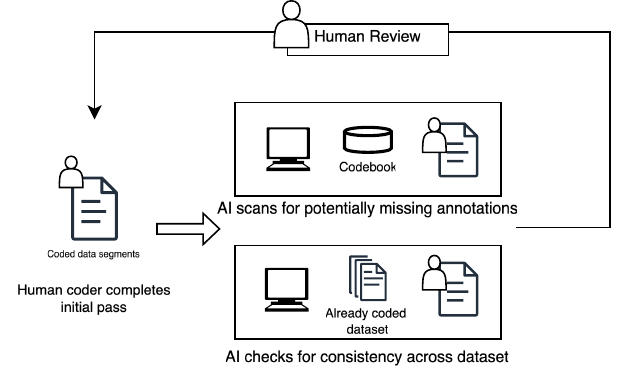}
\par
\caption{[AI for Validation] \textnormal{AI helps identify errors and inconsistencies while keeping researchers actively engaged, e.g., by suggesting potentially missing annotations and providing them for review.}}
\Description{Figure 6 visualizes the role AI for Validation. It depicts a small flowchart with 4 elements. 
The start point is labeled Human coder completes initial task. It shows a pictogram of a person with a pictogram of a paper sheet, that is labeled Coded data segments. From here, an arrow leads to 2 elements representing different possible scenarios. Element 1 contains a pictogram of a computer, a pictogram of a codebook, and a pictogram of a person with a paper sheet. It is labeled AI scans for potentially missing annotations. Element 2 contains a symbol for a computer, a pictogram of 3 paper sheets labeled as Already coded data set, and a pictogram of person with a paper sheet. It is labeled AI checks for consistency across dataset.}
\label{fig:ai_validation}
\end{figure}

\xhdr{AI for Refinement \& Explanations}
Once researchers have developed an initial understanding of the data, AI may assist with refining codebooks by suggesting clearer names, definitions, examples, or restructuring.
For example, P2 noted that AI might suggest merging or grouping similar codes to increase clarity.
For instance, MAXQDA's AI features\footnote{https://www.maxqda.com/blogpost/ai-coding-of-qualitative-data} provide explanations for coded segments, aiding researchers in organizing and refining their codebooks.

\begin{figure}[!htbp]
\centering
\includegraphics[width=.9\linewidth]{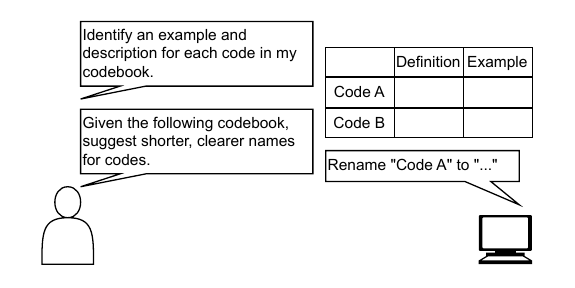}
\par
\caption{[AI for Refinement \& Explanations] \textnormal{AI provides support by offering concise code definitions and names, e.g., for the final codebook, and suitable quotes for codes.}}
\Description{Figure 7 visualizes the role AI for Refinement and Explanations. It depicts pictogram of a person, and a pictogram of a computer, conversing with each other. The conversation has 3 elements with speech bubbles for each. 
In the first element, the person asks: Given the following codebook and dataset, provide descriptions and examples for each code. The speech bubble depicts a book and a dataset. The computer answers with a codetable, consisting of 2 rows for Code A, Code B, and 2 columns titled Definition and Example. 
In the second element, the person asks: Given the following codebook and database, suggest shorter, clearer names for codes. The speech bubble depicts a book and a dataset. The computer answers with Code A, and an arrow leading to 3 dots to represent alternatives. 
In the third element, the person asks: Suggest codes that support the rational of this code. The speech bubble depicts a book and a dataset. The computer answers: Quote for A: ‘...’.}
\label{fig:ai_reasoning}
\end{figure}

\xhdr{AI for Analytics}
Participants also described AI providing reflexive, data-driven insights, such as warnings about overly granular codebooks, expected reliability issues, or frequency patterns.
P12 suggested an AI tool warning them about the codebook getting too long.
P9 suggested a table that shows quantitative statistics on the frequency and patterns of each code, which can be used for further analysis, without taking it away from them.
Taken together, suggestions can support researchers \textit{reflexively}: instead of generating reasons and justifications, they help researchers develop their own.
For instance, Annota~\cite{paleaAnnotaPeerbasedAI2024} marks passages for potential additional improvements.

\begin{figure}[!htbp]
\centering
\includegraphics[width=.8\linewidth]{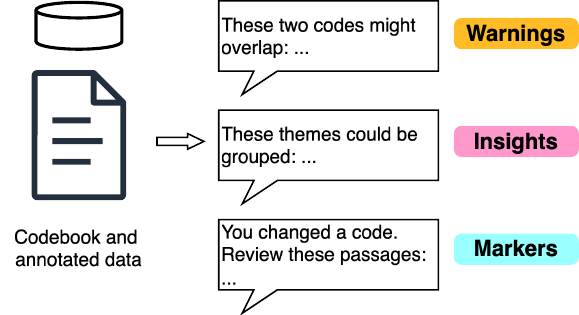}
\par
\caption{[AI for Analytics] \textnormal{AI identifies patterns in coding and provides warnings to support decision-making, e.g., when codes overlap or themes could be grouped.}}
\Description{Figure 8 visualizes the role AI for Analytics. It depicts a small flowchart with 2 steps. 
Step 1 depicts a Codebook and annotated data, represented by pictograms of a paper sheet and a cylinder. An arrow leads to 3 speech bubbles. 
The first bubble says: These two codes might overlap: .... It is labeled Warnings. 
The second bubble says: These themes could be grouped: ... . It is labeled Insights. 
The third bubble says: You changed a code. Review these passages: ... . It is labeled Markers.}
\label{fig:ai_analytics}
\end{figure}

\xhdr{AI Suggestions}
AI suggests codes based on its analysis, which researchers can accept, modify, or reject, maintaining control over the process. 
Our findings indicate that participants were generally open to receiving AI recommendations for new themes and suggestions on where to apply a code.
P8 ideated that \textit{``the tool suggests an important code, or a frequently mentioned word.''}
This openness was particularly evident among participants working with large datasets or repetitive coding tasks.
Existing tools exemplify this role~\cite{maratheSemiAutomatedCodingQualitative2018,rietzCodyAIBasedSystem2021,gao2024collabcoder}.

\begin{figure}[!htbp]
\centering
\includegraphics[width=.7\linewidth]{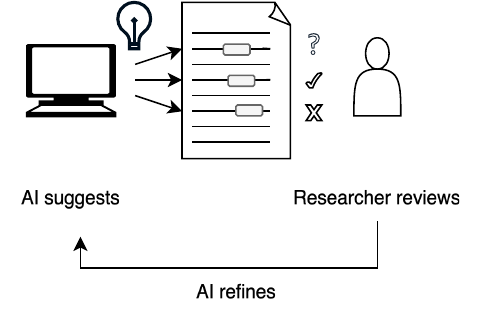}
\par
\caption{[AI Suggestions] \textnormal{AI analyzes the data and recommends codes which can be reviewed by the researcher and refined by the AI, preserving human agency.}}
\Description{Figure 9 visualizes the role AI Suggestions. It depicts a small flowchart with 3 elements. 
Element 1 is labeled AI suggests. It depicts a computer with a light bulb. 3 arrows point to element 2, a paper sheet with lines and 3 squares representing highlights. 
Element 3 is labeled Researcher reviews. It depicts a pictogram of a person, and a question mark, a checkmark, and an x next to the paper sheet. From here, an arrow leads back to element 1. It is labeled AI refines.}
\label{fig:ai_suggestions}
\end{figure}

\xhdr{AI as a Sparring Partner}
Some described AI as a ``sparring partner'', a conversational collaborator that proposes alternatives, challenges assumptions, and supports sensemaking throughout the workflow.
Participants even described this as supporting the existing human collaboration as a third instance or sparring partner (P7; P6; P4), with AI sometimes mediating consensus work (also see \textit{AI as a Mediator}).
Notably, participants preferred an interactive or conversation-like AI (P7). 
For instance, Chew et al.~\cite{chew2023llm} propose a holistic role for \acp{LLM} in co-developing and revising codebooks and conducting final coding.

\begin{figure}[!htbp]
\centering
\includegraphics[width=.99\linewidth]{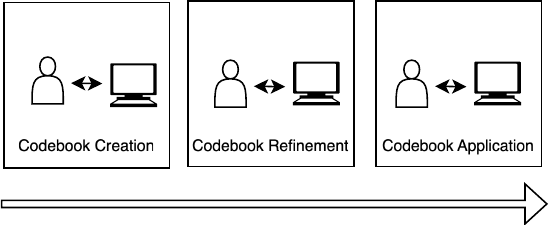}
\par
\caption{[AI as a Sparring Partner] \textnormal{Researchers and AI collaborate interactively throughout the \ac{QDA} workflow, from codebook creation through refinement to application.}}
\Description{Figure 10 visualizes the role AI as a Sparring Partner. It contains 3 elements with a progress arrow underneath. The 3 elements each depict 2 pictograms: a person, a computer, and a double arrow between them.
The first element is labeled Codebook Creation. 
The second element is labeled Codebook Refinement. 
The third element is labeled Codebook Application.}
\label{fig:ai_conversation}
\end{figure}

\subsection{\textbf{High AI Involvement}: AI-Led Workflows}
At this level, AI takes the lead in the \ac{QDA} process, performing tasks autonomously with varying degrees of human oversight.
However, participants raised ethical concerns, including the displacement of researchers and the risk of over-reliance on automated interpretation.
These tensions underscore that full AI delegation remains contentious and highly context-dependent.

\xhdr{AI with Human-in-the-Loop}
In these configurations, AI performs core tasks such as initial coding or codebook generation, while researchers iteratively refine outputs through feedback.
As P8 noted, \textit{``The AI suggests a codebook, and I give feedback that the tool uses to make an updated suggestion.''}
\citet{cohnHumanLoopLLMApproach2024} propose a human-in-the-loop prompt engineering approach to collaborative discourse analysis.
\citet{daiLLMloopLeveragingLarge2023a} suggest an LLM-in-the-loop framework to develop an initial codebook.
This differs from the AI Suggestions role (\autoref{subsec:moderate}) in that here, AI drives the initial output and iteratively incorporates researcher feedback, whereas in the suggestion role, the researcher leads the analysis and AI contributes individual recommendations.

\begin{figure}[!htbp]
\centering
\includegraphics[width=.7\linewidth]{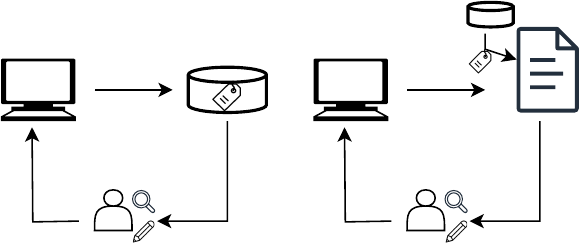}
\par
\caption{[AI with Human-in-the-Loop] \textnormal{AI takes the lead in \ac{QDA}, \eg, creating a codebook draft (left) or performing initial coding (right), with human oversight throughout.}}
\Description{Figure 11 visualizes the role AI with Human-in-the-Loop. It depicts 2 small flowcharts, visualizing 2 different possible workflows, each consisting of 3 elements. 
The first depicts pictogram of a computer, with an arrow leading to a cylinder including a tag, representing a codebook draft. An arrow leads to a pictogram of a person with a magnifier symbol and a pen symbol. From here, an arrow leads back to the computer pictogram.
The second depicts pictogram of a computer, with an arrow leading to a pictogram of a paper sheet. A cylinder with an arrow and a tag leads to the paper sheet, representing initial coding. An arrow leads to a pictogram of a person with a magnifier symbol and a pen symbol. From here, an arrow leads back to the computer pictogram.}
\label{fig:ai_loop}
\end{figure}

\xhdr{AI with Human Approval}
Other participants envisioned workflows where AI acts autonomously but requires human approval, e.g., reviewing random subsamples.
P13 suggested, \textit{``AI could apply my codebook, and I manually review a random subsample.''}
Recent studies use LLMs as additional coders that are evaluated in a final step~\cite{taiExaminationUseLarge2024, banoAIHumanReasoning2023, xiaoSupportingQualitativeAnalysis2023,drapalUsingLargeLanguage2023}.

\begin{figure}[!htbp]
\centering
\includegraphics[width=.6\linewidth]{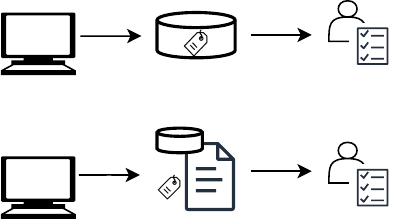}
\par
\caption{[AI with Human Approval] \textnormal{AI creates a codebook draft (top) or assigns codes (bottom). A researcher approves before decisions are finalized and changes are applied.}}
\Description{Figure 12 visualizes the role AI with Human Approval. It depicts 2 small flowcharts, visualizing 2 different possible workflows, each consisting of 3 elements. 
The first depicts computer, with an arrow leading to a cylinder containing a tag, representing a codebook draft. From here, an arrow leads to a pictogram of a person with a paper sheet depicting 3 lines for text and 3 checkmarks, representing researcher approval. 
The second depicts a computer, with an arrow leading to a paper sheet with a cylinder and a tag next to it, representing code assignment. From here, an arrow leads to a pictogram of a person with a paper sheet depicting 3 lines for text and 3 checkmarks, representing researcher approval.}
\label{fig:ai_approval}
\end{figure}

\xhdr{AI with Conditional Autonomy}
AI operates independently in defined scenarios and refers more challenging cases to human researchers.
P4 noted, \textit{``I see AI as tooling for low-hanging fruit, so I can focus on more complex codes.''}
For example, AQUA~\cite{lennonDevelopingTestingAutomated2021} suggests applying AI coding for codes where AI-human IRR matches inter-human agreement.

\begin{figure}[!htbp]
\centering
\includegraphics[width=.7\linewidth]{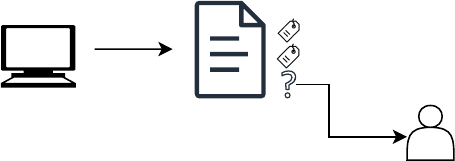}
\par
\caption{[AI with Conditional Autonomy] \textnormal{AI autonomously handles routine coding but alerts researchers when it encounters data requiring nuanced understanding or deviating from the specified criteria.}}
\label{fig:ai_conditional}
\Description{Figure 13 visualizes the role AI with Conditional Autonomy. It depicts a small flowchart consisting of 3 elements. 
Element 1 depicts a computer. An arrow leads to element 2, a pictogram of a paper sheet with two tags and a question mark next to it. From the question mark, an arrow leads to element 3, a pictogram of a person.}
\end{figure}

\xhdr{Full or near-full AI Delegation}
While no participant supported fully autonomous QDA end-to-end, some were open to delegating individual steps, such as applying a finalized codebook or generating an initial one (P8).
Agentic AI systems and multi-agent approaches illustrate this direction~\cite{rasheedCanLargeLanguage2024}. However, participants emphasized that such use cases demand heightened scrutiny due to risks around validity, reflexivity, and loss of human agency~\cite{jiangSupportingSerendipityOpportunities2021,feustonPuttingToolsTheir2021}.
OpenAI's Viable can be used autonomously for analysis of unstructured data at scale, for example, in businesses that require regular customer feedback evaluations~\cite{AccuratelyAnalyzingLarge2024}.

\subsection{Summary and Implications}
While \ac{HCI} researchers were open to exploring the potential of AI support for \ac{QDA}, they emphasized the need for responsible design that respects research ethics, quality, and researcher control.
Our framework highlights that responsible AI integration in QDA is less about replacing human analysis and collaboration, and more about negotiating agency, trust, and epistemic responsibility. 
Notably, many participants articulated multiple ideas across different levels of integration, underscoring that acceptable AI involvement is fluid and context-dependent rather than fixed.

We stress that this framework is primarily descriptive rather than prescriptive: it maps the design space that participants articulated, without recommending specific levels of AI involvement for specific contexts.
Our positionality (early-career, WEIRD HCI researchers with experience in qualitative methods and AI) shapes this synthesis.
Open questions remain about who benefits (or is harmed), what labor is displaced or amplified, and whether productivity should be the dominant metric guiding adoption.
Our findings suggest that a key driving factor for AI adoption was to increase productivity, raising difficult questions around whether speed should be a primary goal in academic qualitative research.
If AI integration is primarily motivated by efficiency, it risks devaluing the slow, iterative processes through which qualitative researchers develop deep understanding of their data~\cite{jiangSupportingSerendipityOpportunities2021,paulus2024minutes}, particularly in institutional contexts where publication pressure may make AI-driven acceleration attractive for structural rather than methodological reasons.

At the same time, we do not view the framework as static. As new interaction paradigms emerge, such as more conversational, adaptive, or team-aware AI systems, and as collaborative research practices evolve, the roles and boundaries we identify will likely shift.
We therefore expect the framework to be iteratively refined through future empirical work, serving as a living scaffold that evolves alongside changing human–AI relationships in qualitative research.

Overall, our findings underscore that AI support in QDA is neither neutral nor inevitable. Any integration must carefully weigh ethical, methodological, and practical considerations.

\section{How our Framework Might be Used}

Our framework aims to empower HCI researchers to critically examine if, where, and how to incorporate AI tools into QDA.
Rather than defaulting to AI-driven coding, it highlights multiple ways AI could support QDA while preserving research autonomy. 
Determining the appropriate role of AI requires careful consideration of research goals, data types, and \ac{QDA} methods. Below, we outline key considerations for using the framework to decide what role AI should play in a given study.

\xhdr{Determining the appropriate level of AI involvement}
AI involvement in QDA is highly context-dependent and should thus not be treated as default.
Researchers must first decide whether AI is appropriate at all, or whether other technologies could support the task. This requires literacy in both AI and QDA, particularly given risks of overreliance and explainability pitfalls~\cite{feustonPuttingToolsTheir2021, okolo2022making, Buccinca2021}. Such decisions should be made collaboratively within research teams.
For example, large-scale or structured datasets may benefit from AI’s ability to handle pattern recognition, while nuanced or sensitive data might necessitate limited AI involvement to maintain interpretive depth and ethical considerations. 
Research goals -- whether they focus on exploration, theory generation, or confirmation -- also dictate how AI can best support \ac{QDA}.
AI may be more useful in early stages (e.g., surfacing patterns) or later stages (e.g., consistency checks), while in some cases it may best support technical tasks rather than analysis itself.

\xhdr{Deciding when AI should lead or support}
Related work shows that coding performance can depend on various factors of complexity, \eg, the required level of judgment and interpretation ~\cite{ortloffDifferentResearchersDifferent2023,kirsten2024decoding}.
Routine tasks such as transcription may warrant greater AI autonomy, whereas ambiguous or interpretive tasks call for stronger human leadership.
In addition, researchers' comfort and familiarity with AI tools as well as \ac{QDA} should guide their level of involvement~\cite{lubars2019ask, gauthier2022computational}.
Design strategies such as restricting the visibility of AI outputs or displaying performance disclaimers may further support appropriate use~\cite{bhatt2024should, lam2024concept}.

\xhdr{Identifying stages where human oversight is essential}
Human oversight remains critical at key points in the workflow.
Researchers should determine at which stages oversight is crucial -- such as after each coding pass, document review, or when salient patterns emerge.
These checkpoints support collaborative sensemaking and ownership rather than replacing human analysis.
When AI operates conditionally or autonomously, researchers should define clear triggers for human intervention to preserve agency.

\xhdr{Deciding on the degree of human control}
Researchers must balance AI efficiency with the ability to intervene and customize outputs~\cite{wang2020human, morris2024human}. AI tools should allow manual refinement of codes or summaries to align with researchers’ interpretive perspectives.
However, excessive oversight can introduce new burdens~\cite{feustonPuttingToolsTheir2021, lubars2019ask, gauthier2022computational, lam2024concept}. In such cases, less AI involvement may be preferable.

\xhdr{AI integration across theoretical QDA frameworks}
Our findings suggest that researchers may not rigidly adhere to a single QDA framework in practice.
AI systems tightly coupled to specific methods may therefore be counterproductive. Flexible tools that accommodate evolving workflows are likely more useful.
However, projects grounded in specific frameworks may benefit from adjustable or specialized AI models, for example by instructing \acp{LLM} to adopt a given theoretical lens.
Furthermore, examining how AI-generated codebooks differ when different frameworks are applied may provide researchers with new insights into their data, supporting more informed decisions about the most suitable coding framework for their project.
When AI is used as a supporting agent rather than the primary coder, its suggestions must align with the theoretical and methodological context defined by human coders.
Misaligned suggestions can disrupt the research process and frustrate researchers by introducing inconsistencies and workload. %
Ensuring compatibility between AI tools and the methodological requirements of a study is therefore crucial for maintaining coherence and usability in QDA workflows.

\xhdr{Evaluating Ethical Implications and Compliance}
AI integration in QDA raises ethical concerns around privacy, security, consent, and intellectual ownership. Participants should be informed when AI is used in analysis, and safeguards must be in place when sharing data with AI systems.
Finally, depending on the level of AI autonomy, a normative discussion on violations of intellectual integrity and ownership is necessary.
For example, it is necessary to report the level of AI usage in research outputs to maintain transparency and accountability in the research process, especially in the face of issues with transparency and reproducibility~\cite{wacharamanotham2020transparency, fiesler2019qualitative}.

\xhdr{Call to Action} %
Researchers already use AI throughout their research process (\eg~\cite{kapania2025categorizing, argyle2023out, hamalainen2023evaluating, oppenlaender2023mapping}), and popular QDA software increasingly integrates AI features.
Our findings show that researchers do not seek a single \emph{best} way to integrate AI into QDA. Instead, they emphasize choice, context, and control.
This suggests that the key challenge for AI-assisted QDA is not increasing autonomy or speed, but supporting deliberate, situated decisions about whether and how AI should be involved at all.

\section{Discussion}
\label{qdallm::sec::discussion}

Our findings demonstrate that integrating AI into \ac{QDA} involves more than just identifying which analytic tasks can be automated.
It also requires aligning AI support with situated, collaborative, and methodologically diverse research practices. 
In this section, we discuss how our results build on prior work on AI-supported \ac{QDA}, highlight tensions between efficiency, researcher agency, and interpretive depth, and reflect on broader implications.

\subsection{Synthesizing Prior Work on AI-Supported QDA}
\label{subsec:synthesizing-prior-work-on-ai-supported-qda}
Our findings extend prior HCI research on AI-supported \acf{QDA}~\cite{feustonPuttingToolsTheir2021, jiangSupportingSerendipityOpportunities2021, maratheSemiAutomatedCodingQualitative2018, gaoCoAIcoderExaminingEffectiveness2023, kapania2025categorizing, zhangRedefiningQualitativeAnalysis2024, chatzichristos2025qualitative} by grounding this body of work in an empirical account of how researchers conduct \ac{QDA} in practice.  
The diversity of \ac{QDA} approaches, persistent skepticism toward AI integration~\cite{jiangSupportingSerendipityOpportunities2021, feustonPuttingToolsTheir2021, chatzichristos2025qualitative}, varied epistemic backgrounds of HCI researchers, and the fundamentally social and collaborative nature of \ac{QDA}~\cite{fiesler2019qualitative, mcdonald2019reliability, wallace2017technologies, gauthier2022computational} underscore the need for careful and context-sensitive integration of AI into these workflows.

\xhdr{Workflows and Situated \ac{QDA} Practice}
While earlier work examined tensions around reliability, transparency, and reflexivity in qualitative research~\cite{fiesler2019qualitative,mcdonald2019reliability}, these conversations remain fragmented across subfields and methodological traditions. 
By documenting researchers’ workflows in detail, we synthesize these debates into an empirically grounded overview that makes visible recurring pain points -- such as negotiating consensus or applying IRR -- alongside the rewarding aspects of collaborative sensemaking.
This workflow-centered account situates AI-supported \ac{QDA} within HCI's long-standing recognition of qualitative analysis as a negotiated and social process, as documented across HCI venues~\cite{wallace2017technologies}. 
Building on calls for flexible and customizable AI designs for \ac{QDA}~\cite{feustonPuttingToolsTheir2021}, we extend prior work to the era of generative AI, highlighting emerging concerns such as hallucinations and shifting researcher expectations around AI assistance.

\xhdr{Shifting Attitudes toward AI Collaboration}
\citet{jiangSupportingSerendipityOpportunities2021} applied the AI task delegability framework~\cite{lubars2019ask} to assess which \ac{QDA} tasks AI should perform, considering factors like motivation, risk, trust, and difficulty. 
In line with research outside HCI~\cite{dellafiore2025artificial}, they found researchers to be largely cautious of AI intervention.
In contrast, we observe growing openness to AI-based support, particularly when human agency and control are preserved. 
Recent work further identified a generational divide in openness, with early career researchers being more open, and more experienced researchers being more cautious~\cite{chatzichristos2025qualitative}. 
These patterns highlight the need for careful design that accommodates varying proficiencies in AI capability as well as \ac{QDA}, reminding researchers of the limitations of AI, and allowing for evaluation and fine-tuning while avoiding excessive workload of doing so~\cite{feustonPuttingToolsTheir2021, lubars2019ask, gauthier2022computational, lam2024concept}. 
However, the field of Human-Centered Explainable AI is still in the making~\cite{ehsan2024human}, and research shows deficiencies in meeting user needs and exacerbating issues of algorithmic opacity~\cite{ehsan2021operationalizing, ehsan2022human, ehsan2023charting, khalili2023against}. Okolo~\cite{okolo2022making} further found work on Explainable AI to be centered around the Global North, with concerning gaps in knowledge about the Global Souths, warning about ``explainability pitfalls''~\cite{ehsan2024explainability} causing overreliance on AI.
Crucially, integrating AI into QDA introduces a shift from Human-AI interaction to Human-AI collaboration that involves ``mutual goal understanding, preemptive task co-management and shared progress tracking''~\cite{wang2020human} in sensemaking, and more research is needed to determine how this can be best supported to avoid lopsided collaboration structures in AI-assisted \ac{QDA}~\cite{morris2024human, wang2020human}. Related work emphasizes this collaboration needs to be highly interactive, facilitated by mutual learning that fosters a symbiotic partnership that preserves researcher autonomy, rather than lending autonomy to AI~\cite{feustonPuttingToolsTheir2021, cho2020role,nagao2019symbiosis}. 
Our results have shown that collaborative AI roles were indeed perceived as favorable \ac{QDA} scenarios by our study participants.

\xhdr{Messiness, Productivity, and the Limits of Automation}
Our findings align with prior insights on the messiness and uncertainty of data analysis, which our participants also acknowledged.
However, our interviews reveal that not all pain points are areas where researchers desire automation; for example, participants often prefer to engage directly with the ``messiness'' of their data to better understand it~\cite{paulus2024minutes, jiangSupportingSerendipityOpportunities2021}. 
Still, AI can support engagement with messiness at scale, bridging ``small'' and ``big'' data analysis approaches~\cite{jiangSupportingSerendipityOpportunities2021, lam2024concept, baumer2017comparing, muller2016machine}.

Consistent with \citet{jiangSupportingSerendipityOpportunities2021}, our participants emphasized the intrinsic value they find within the \ac{QDA} process, including the satisfaction of uncovering patterns and learning through collaboration, which is particularly relevant in HCI, a field that comprises researchers with various backgrounds and epistemic traditions~\cite{jiangSupportingSerendipityOpportunities2021, mcdonald2019reliability, bjorn2015multiple, fiesler2019qualitative}. 
These insights challenge productivity-focused narratives surrounding AI, highlighting potential tensions between efficiency and \ac{QDA}’s interpretive depth~\cite{chatzichristos2025qualitative, paulus2024minutes}, social dynamics, and opportunities for learning, skill-building, and knowledge transmission~\cite{jiangSupportingSerendipityOpportunities2021}.

\subsection{Tensions, Values, and Inequities in AI-Supported QDA}
While AI-supported QDA offers novel opportunities, concerns about output quality and bias remain central.  
These concerns are particularly salient when working with non-English data~\cite{heseltine2024large, suter2024using}, engaging minority communities, or conducting research in the Global South, as most AI systems are trained on English and other high-resource languages and are not ``culturally agnostic''~\cite{paulus2024minutes}. 
As a result, such systems often privilege dominant or high-frequency representations~\cite{feustonPuttingToolsTheir2021}, flatten linguistic nuance, and embed familiar Western styles of reasoning and analysis, making it harder to account for alternative narrative forms, cultural contexts, or theoretical traditions. 
Given that perspectives from the Global Souths and majority world are already marginalized within HCI~\cite{kumar2021braving, sturm2015weird, linxen2021weird, hasegawa2024weird}, AI-supported \ac{QDA} risks further amplifying these imbalances by embedding dominant interpretative frameworks and overlooking locally situated meanings~\cite{kumar2021braving}. 
Thus, it is essential to build culturally sensitive and context-aware approaches to AI-supported \ac{QDA}. 

Finally, the integration of AI into \ac{QDA} unfolds alongside broader debates in HCI concerning citational justice~\cite{kumar2021braving}, transparency of research practices~\cite{wallace2017technologies}, and epistemic tensions~\cite{kumar2021braving, fiesler2019qualitative, pierre2021getting, ajmani2024whose, soden2024evaluating}. 
Researchers caution against a ``gradual return to a methodological positivism''~\cite{chatzichristos2025qualitative} that AI could facilitate within interpretive \ac{QDA}, while at the same time emphasizing the complementary potential of computational, positivist, and interpretive approaches~\cite{baumer2017comparing, chatzichristos2025qualitative, gauthier2022computational, lam2024concept, muller2016machine}. 
Situated within broader politics of knowledge production in HCI~\cite{kumar2021braving, collective2021following, soden2024evaluating}, these dynamics are further shaped by uneven access to AI-enabled \ac{QDA} tools, which often require reliable internet access, computational resources, or paid subscriptions -- advantages that disproportionately accrue to well-resourced institutions and researchers.
These tensions call for sustained reflection on epistemic justice and geographically shaped access to resources~\cite{chatzichristos2025qualitative, ajmani2024whose, kumar2021braving}, including tools, time, and funding, to avoid reinforcing existing inequalities through AI integration into \ac{QDA}.

\subsection{Limitations \& Future Work} 

Our study faces several limitations, primarily those common to interview-based research, including self-report, recall, and social desirability biases.
Further studies could benefit from alternative methods, such as observation, screen recordings, diary studies, and subsequent quantitative approaches to measure findings at scale.

Our sample consisted of participants with varying levels of experience in qualitative coding and different data types, which limited our ability to provide a unified, in-depth account of one specific \ac{QDA} approach. 
Still, the diversity of reported \ac{QDA} practices highlights the complexity of AI integration. 
Additionally, our sample is slightly skewed towards early career researchers, and all participants were from HCI research, many also specialized in security and privacy.
This sampling choice reflects the scope of our study: we aimed to understand QDA workflows and AI-support expectations within HCI.
At the same time, this scope likely shaped our findings.
HCI researchers may be more familiar with emerging technologies and more open to computational support than researchers in disciplines where qualitative methods are more established as core methodological traditions.
Our findings therefore may not generalize to other fields where QDA training, epistemological commitments, standards, and attitudes toward automation may differ, and future work should investigate a broader range of disciplinary contexts.
We did not collect data on prior exposure to AI or LLMs, but recognize that participants' related motivational factors~\cite{Bandura.2012}, their differential attitudes~\cite{Ajzen.2019}, even their preconceptions or misconceptions~\cite{Gawronski.2012, Hammer.1996}, and incidental knowledge~\cite{Ferguson.2024} concerning AI tools, can influence their responses. 
Nonetheless, our sample provided detailed insights into the potential for adaptive AI tools that can cater to both novice guidance and expert customization.
Due to our sampling method choice (particularly relying on snowball sampling through our professional networks), our sample consisted of people working in a WEIRD context. This inherently limits the generalizability of our findings as well as how globally representative they are. Future work can and should interview researchers from other contexts to bring in additional perspectives regarding how AI should, or should not, be involved in QDA. 

While our framework outlines numerous ways AI tools could be integrated into \ac{QDA}, we do not empirically evaluate how these integrations could be implemented or how they impact the overall \ac{QDA} process.
Future research should assess the performance and usability of these \ac{AI} tools in \ac{QDA}, exploring how effectively they support researchers and whether they facilitate the analysis in meaningful ways.
We also stress the importance of further research on the effects of AI on human decision-making in \ac{QDA}. 
Studies have shown that users tend to over-rely on \ac{AI} recommendations, even in low-involvement scenarios, highlighting the risk of blind acceptance and overreliance~\cite{Buccinca2021}.

\section{Conclusion}
\label{qdallm::sec::conclusion}

In this paper, we investigated how \acf{AI}, particularly \acfp{LLM}, can be integrated into \acf{QDA} workflows from the perspective of \ac{HCI} researchers.
Our interview study with 15 researchers experienced in \ac{QDA} revealed the varied and context-dependent nature of real-life \ac{QDA} practices.
Most participants expressed openness to greater \ac{AI} involvement in their \ac{QDA} workflows, indicating a notable shift from attitudes observed in studies conducted prior to the advent of \acp{LLM}.
They were interested in AI tools that can assist with tasks such as coding, summarizing, or facilitating consensus among researchers, while emphasizing the importance of maintaining human oversight. 
Ethical concerns, including data privacy and the reliability of \ac{AI} outputs, were prominent, underscoring the need for \ac{AI} tools to be transparent, dependable, and aligned with established research standards.
In response to these needs, we propose a framework categorizing \ac{AI} involvement in \ac{QDA} into three levels: 
minimal, moderate, and high. 
This framework allows researchers to select the appropriate level of \ac{AI} assistance based on their specific project requirements and context, ensuring that \ac{AI} complements rather than replaces human judgment.
In doing so, our work contributes to more reliable and ethical research practices that benefit the broader academic and societal landscape.
We hope this work contributes to an ongoing conversation within the \ac{HCI} community about designing \ac{AI} tools that are not only technically capable but also sensitive to the human-centered, collaborative, and interpretive values that underpin qualitative inquiry.

\begin{acks}
This work was supported by the \href{https://rc-trust.ai}{Research Center Trustworthy Data Science and Security}, one of the Research Alliance Centers within the \href{https://uaruhr.de}{UA Ruhr}.
\end{acks}

\section*{Use of AI Tools}
\acp{LLM} were used to assist with grammar and spelling corrections during the preparation of this manuscript. All generated outputs were reviewed and verified by the authors.

\balance
\bibliographystyle{ACM-Reference-Format}
\bibliography{bibliography}

\newpage
\onecolumn
\appendix
\section{Detailed Interview Guide}
\label{qdallm::sec::appendix0}
Below, we provide the detailed interview guide for our study.

\subsection*{Welcome \& Study Information}
\begin{itemize}[leftmargin=1cm]
    \item Welcome the participant and briefly introduce ourselves.
    \item Check the informed consent form.
    \item Provide information on the study and data handling:
    \begin{itemize}
        \item Participation is voluntary and can be discontinued or withdrawn at any time.
        \item This is an exploratory interview, which will last no longer than 45 minutes.
        \item There is no compensation for participation.
        \item We collect some personal information for our scientific research, which will be stored in pseudonymized form.
        \item Recorded data is subject to the guidelines of the General Data Protection Regulation~(GDPR).
    \end{itemize}
    \item Ask the participant if they have any questions.
    \item Explain the purpose and goal of this research:
    \begin{itemize}
        \item The purpose of this study is to understand the qualitative coding practices and challenges of HCI researchers, as well as expectations for computer-assisted qualitative coding.
        \item Your participation will help us achieve this research goal.
        \item Results may be presented in a scientific publication.
    \end{itemize}
    \item Inform the participant about the study procedure:
    \begin{itemize}
        \item Ask if the participant completed the online survey.
        \item Next: Interview session on qualitative research experiences, challenges, and anticipations, divided into three parts.
        \item We will take notes, and we will ask you to sketch your thoughts on paper.
        \item We would like to collect your paper sketches afterwards.
    \end{itemize}
    \item Ask the participant if they have any questions.
    \item Emphasize that there is no pressure: Workflows do not have to be textbook standard, and there is no judgment.
\end{itemize}

\subsection*{Interview Session on Qualitative Research Experiences}
\begin{itemize}[leftmargin=1cm]
    \item \textbf{Q1: Can you give us a high-level overview of your research process from data collection to analysis? Please use this paper sheet to sketch your workflow and explain what you are sketching. You can take your time in sketching and explain afterwards if you prefer.}
    \begin{itemize}
        \item Please recall your last research project and describe how you coded and analyzed the data.
        \item What kind of data (data type) did you code?
        \item Did you use any additional resources when coding?
        \item What tools and methodology did you use?
        \item How many people are usually involved in this process? How do you collaborate?
    \end{itemize}
    \item \textbf{Q2: What parts of coding do you find the most interesting? Which parts the most tedious? You can use your paper sketch to talk about individual steps and challenges.}
    \begin{itemize}
        \item Which parts do you personally like the most and why?
        \item What are your pain points when it comes to qualitative analysis? What do you dislike about these parts?
        \begin{itemize}
            \item At which stage do you spend the most time?
            \item Which task/step do you enjoy doing the least and why? What are your major challenges/difficulties when coding?
            \item To give ideas: segmenting, collaborating, iterating \& revising, refining codes, etc.
        \end{itemize}
        \item What features do you like and dislike most about the QDA tools you are using?
        \item Where do your QDA tools support you?
    \end{itemize}
    \item \textbf{Q3: What are your expectations for AI-assisted QDA? Under which conditions and at which steps are you willing to use AI technologies for QDA?}
    \begin{itemize}
        \item Would you be willing to use AI software that partially automates coding? Under which conditions?
        \item Where in your research process would you see such an assistive tool?
        \item Are there any parts of the QDA process you wish were automated? Are there parts you think should absolutely not be automated?
        \item Looking at emergent technologies (think of ChatGPT, GPT-4), how would you envision these Large Language Models in the research process? (Thoughts, concerns)
        \item Optionally: What would such a tool look like for you?
    \end{itemize}
\end{itemize}

\subsection*{Answering Questions \& Farewell}
\begin{itemize}[leftmargin=1cm]
    \item Is there anything you would like to add? Do you have any questions?
    \item Thank you for your participation!
\end{itemize}

\newpage
\section{Complete Codebooks from Our Interviews}
\label{qdallm::sec::appendix1}

\begin{table*}[htb]
\small
\centering
\caption{Codebook for \textit{Tools and Resources} based on our interviews with 15 participants. The codebook includes all codes, along with their descriptions, examples, and the number of participants per code.}
\label{qdallm::tab::codes_tools}

\setlength{\tabcolsep}{4pt}
\renewcommand{\arraystretch}{1.08}

\begin{tabular}{@{}p{2.5cm}p{6cm}p{6cm}r@{}}
\toprule
\textbf{Code} & \textbf{Description} & \textbf{Examples} & \textbf{Qty.} \\
\midrule

\textit{Google Docs} & cloud-based application to collaborate on text-based documents in real-time from any device & ``Google Docs (nice collaboration features)'' & 1 \\

\textit{Google Drive} & cloud-based storage service to store, share, and access files from any device & ``Google Drive which is very convenient for sharing'' & 1 \\

\textit{Google Sheets} & cloud-based application to collaborate on spreadsheets in real-time from any device & ``enter data into a Google Sheet and analyze it'' & 2 \\

\textit{Microsoft Word} & word-processing software to create, edit, and share text-based documents & ``Shared Word document: as a shared memo'' & 2 \\

\textit{Microsoft Excel} & software to organize, analyze, and visualize data using spreadsheets & ``Codes merged into large Excel table'' & 9 \\

\textit{Text Editor} & software to create, edit, and manage plain text files using basic formatting and coding features & ``Text editor: build hierarchies, move codes [\dots], take notes'' & 1 \\

\textit{Memos} & brief written messages for internal communication to convey information, instructions, or reminders & ``make sure each code has a memo'' & 1 \\

\textit{Transcription Tools} & software to convert speech into written text to transcribe audio or video recordings & ``Interviews are conducted, then transcribed'' & 4 \\

\textit{NVivo} & qualitative analysis tool to analyze and visualize data from interviews, surveys, and social media & ``Get the whole website (download) via NVivo'' & 5 \\

\textit{Post-Its} & small, sticky paper notes to write messages that can be attached to surfaces & ``I used to do it manually [...]. I used sticky notes'' & 2 \\

\textit{ATLAS.ti} & qualitative analysis tool for complex text, audio, and video data using coding and visualization & ``ATLAS.ti and NVivo'' & 4 \\

\textit{Dedoose} & web-based application for qualitative and mixed-methods research to analyze and visualize data through coding and collaboration & ``Dedoose for other projects: multiple people can code at the same time'' & 3 \\

\textit{Python} & versatile, high-level programming language, popular for web development and data analysis & ``Python script for data processing'' & 4 \\

\textit{Notes App} & software to create, edit, and organize notes using categorization, search, and cloud synchronization & ``notes app (to upload to iCloud)'' & 1 \\

\textit{MAXQDA} & qualitative data analysis software for systematic organization, coding, and interpretation of text, audio, and video data & ``I used MAXQDA'' & 12 \\

\textit{Miro} & cloud-based platform for teams to brainstorm, plan, and collaborate in real-time on a shared canvas & ``Move to Miro: Online Post-It Platform'' & 2 \\

\textit{Shared Pad} & online collaborative tool to create, edit, and comment on documents in real-time & ``shared pad for shared notes'' & 1 \\

\textit{INCEpTION} & open-source platform for collaborative annotation and knowledge extraction from text, enabling users to manage labeled datasets for machine learning & ``INCEpTION for annotations'' & 1 \\

\textit{API / Crawler} & interface/program to automatically navigate the web to index content and gather data from websites & ``Understanding API/Crawler'' & 1 \\

\bottomrule
\end{tabular}
\end{table*}

\begin{table*}[htb]
\small
\centering
\caption{Codebook for \textit{Data Types} based on our interviews with 15 participants. The codebook includes all codes, along with examples and the number of participants per code.}
\label{qdallm::tab::codes_data}

\setlength{\tabcolsep}{4pt}
\renewcommand{\arraystretch}{1.08}

\begin{tabular}{@{}p{6cm}p{8.0cm}r@{}}
\toprule
\textbf{Code} & \textbf{Examples} & \textbf{Qty.} \\
\midrule

\textit{Audio Recordings} & ``Transcripts of [\dots] audio recordings'' & 2 \\

\textit{Screenshots} & ``I've also done website screenshots'' & 1 \\

\textit{Log Files} & ``software log files to track task performance'' & 1 \\

\textit{Drawing \& Illustrations} & ``I used [\dots] NVivo for coding pictures/images'' & 1 \\

\textit{Social Media Posts} & ``reddit posts'' & 3 \\

\textit{Written Documents} & ``Mostly written data[\dots], no transcription required'' & 2 \\

\textit{Interview Transcripts} & ``Interview Study'' & 8 \\

\textit{Open-Ended Survey Responses} & ``Open-ended survey responses, approximately 2000'' & 1 \\

\textit{Think Aloud Protocols} & ``Think-Aloud Protocols'' & 2 \\

\textit{Screen Recordings} & ``screen capture must be analyzed manually'' & 2 \\

\bottomrule
\end{tabular}
\end{table*}

\begin{table*}[htb]
\small
\centering
\caption{Codebook for \textit{Fun Points} based on our interviews with 15 participants. The codebook includes all codes, along with their descriptions, examples, and the number of participants per code.}
\label{qdallm::tab::codes_fun}

\setlength{\tabcolsep}{4pt}
\renewcommand{\arraystretch}{1.08}

\begin{tabular}{@{}p{3cm}p{6cm}p{6cm}r@{}}
\toprule
\textbf{Code} & \textbf{Description} & \textbf{Examples} & \textbf{Qty.} \\
\midrule

\textit{Interesting Topic} 
& Creating and editing interesting things 
& ``Working on transcripts is fun when the topic is interesting'' 
& 3 \\

\textit{Working With People} 
& Interaction with others, e.g., during interviews 
& ``working with people, lots of contact'' 
& 1 \\

\textit{Coding} 
& Applying the finalized codebook to the collected data sets 
& ``I also really like using the codebook once I have it'' 
& 8 \\

\textit{Finding Interesting Quotes} 
& Extract citations that support the identified themes and the analysis 
& ``I enjoy [\dots] sorting out of interesting quotes'' 
& 3 \\

\textit{Synthesizing Main Themes} 
& Synthesize key points of interest 
& ``When you [\dots] map [themes] to answers to RQs'' 
& 3 \\

\textit{Learnings From Collaboration} 
& Learning from onboarding, collaboration, and training 
& ``Training with Coder 4'' 
& 1 \\

\textit{Mixing Quantitative \& Qualitative Data} 
& Use of mixed methods 
& ``Bringing together mixed methods quantitative and qualitative data'' 
& 1 \\

\textit{Forming The Codebook} 
& Develop and edit the codebook 
& ``Forming the codebook, when we [\dots] see what results we have'' 
& 5 \\

\textit{Organizing \& Finalizing The Codebook} 
& Codebook refinement and finalization 
& ``Break it down into lower level codes'' 
& 1 \\

\textit{Learn \& Get Deeper Insights} 
& Get deeper understanding of a topic 
& ``New experience; [\dots] you get deeper insights into user perspectives'' 
& 4 \\

\textit{Presenting The Results} 
& The presentation of the results, e.g., formulation of the findings 
& ``Analyze and *present* the results'' 
& 2 \\

\textit{Analysis} 
& Data collection analysis 
& ``Analysis is fun'' 
& 4 \\

\textit{Discussion} 
& Interesting discussions with co-workers 
& ``Discussing is fun'' 
& 7 \\

\textit{Familiarizing \& Exploring Data} 
& Familiarisation and exploration of collected data 
& ``getting familiar with the data'' 
& 3 \\

\bottomrule
\end{tabular}
\end{table*}

\begin{table*}[p]
\footnotesize
\centering
\caption{Codebook for \textit{Pain Points} based on our interviews with 15 participants. The codebook includes all codes structured by high-level concepts, along with their descriptions, examples, and the number of participants per code.}
\label{qdallm::tab::codes_pain}

\setlength{\tabcolsep}{3.5pt}
\renewcommand{\arraystretch}{1.02}

\begin{tabular}{@{}p{3.1cm}p{6cm}p{6cm}r@{}}
\toprule
\textbf{Code} & \textbf{Description} & \textbf{Examples} & \textbf{Qty.} \\
\midrule

\multicolumn{4}{@{}l}{\textbf{Codebook Development}} \\

$\hookrightarrow$ \textit{Deciding Focus} 
& Decide on the focus of the analysis and the important aspects of the data 
& ``Decide what to focus on'' 
& 3 \\

$\hookrightarrow$ \textit{Going Through First Data} 
& Browsing the first batch of data to come up with codes and get acquainted 
& ``Data viewing, initial data creation'' 
& 2 \\

$\hookrightarrow$ \textit{Restructuring the Codebook} 
& Restructure the entire codebook if something goes wrong 
& ``Going back and forth when fixing codes'' 
& 8 \\

$\hookrightarrow$ \textit{Developing Hierarchies} 
& Define code hierarchies and relationships, introduce subcodes, etc. 
& ``too many codes, painful to put them into [\dots] themes'' 
& 2 \\

$\hookrightarrow$ \textit{Finding \& Naming Codes} 
& Codes should be named in a way that differentiates and clarifies them 
& ``sometimes codes can get long'' 
& 5 \\

$\hookrightarrow$ \textit{Losing Sight of RQs} 
& Easy to forget what the overarching research goals and RQs are 
& ``what is the RQ? [\dots] easy to lose sight of that'' 
& 1 \\

$\hookrightarrow$ \textit{Overdetailing the Codebook} 
& Codebook is too detailed, too granular, and contains too many codes 
& ``Codebooks become too large and detailed'' 
& 3 \\

$\hookrightarrow$ \textit{Losing Overview of Codes} 
& To lose sight of the mass of codes in the codebook 
& ``It is difficult to keep track of the entire [\dots] codebook.'' 
& 6 \\

\textit{Cleaning Transcriptions} 
& Cleanup of transcripts can be tedious and time-consuming 
& ``Cleaning up the transcripts is very tedious'' 
& 4 \\

\textit{Segmenting} 
& Split interviews into shorter pieces for analysis 
& ``If I have very noisy [transcripts], that can be very difficult'' 
& 2 \\

\textit{Synthesis \& Analysis} 
& Synthesizing and analyzing the data is challenging 
& ``Synthesize: challenging but also exciting and important step'' 
& 1 \\

\midrule
\multicolumn{4}{@{}l}{\textbf{Collaboration}} \\

$\hookrightarrow$ \textit{Relying on Others} 
& Dependence on others may be a problem 
& ``you have to rely on another person, who may fail'' 
& 1 \\

$\hookrightarrow$ \textit{Initial Discussions} 
& The initial discussions about the structure of the codebook 
& ``Hard at the beginning'' 
& 1 \\

$\hookrightarrow$ \textit{Iterative Cycles} 
& Repeating cycle of discussion, agreement, and interpretation of codes 
& ``discussing code about code with the same discussion'' 
& 7 \\

$\hookrightarrow$ \textit{On-boarding Others} 
& Training other coders can be time-consuming 
& ``can be a bit tedious in the beginning, especially the briefing/onboarding.'' 
& 1 \\

$\hookrightarrow$ \textit{Long Discussions} 
& Dialogue and discussion take up a lot of time 
& ``two people with strong opinions need to agree'' 
& 4 \\

$\hookrightarrow$ \textit{Interpretation Discrepancies} 
& Discussing how coders interpret the differences 
& ``difficulties with others when you understand the categories differently'' 
& 9 \\

\midrule
\multicolumn{4}{@{}l}{\textbf{Codebook Application}} \\

$\hookrightarrow$ \textit{Tedious Process} 
& Assigning codes from the codebook to data segments is tedious 
& ``Could also be a bit tedious'' 
& 6 \\

$\hookrightarrow$ \textit{Making Adjustments Later On} 
& To update a code, you have to go back and possibly re-code already coded data 
& ``Going back and forth when fixing codes'' 
& 5 \\

\textit{IRR Issues} 
& Obtaining a good IRR takes time and is complicated 
& ``Discussions, just to get IRR'' 
& 5 \\

\textit{Time-Consuming Process} 
& Coding takes time 
& ``If the coding process takes a long time'' 
& 3 \\

\midrule
\multicolumn{4}{@{}l}{\textbf{Lack of Structure and Guidance}} \\

$\hookrightarrow$ \textit{Need For Assistance} 
& Participant wishes assistance with coding 
& ``A second pair of eyes helps'' 
& 4 \\

$\hookrightarrow$ \textit{Messiness} 
& Data can be complex and confusing 
& ``Social media posts can be very unstructured'' 
& 3 \\

$\hookrightarrow$ \textit{Subjectivity} 
& Interpretation of data can be subjective 
& ``Codes can be subjective'' 
& 1 \\

$\hookrightarrow$ \textit{Uncertainty} 
& Interpretation of data can be ambiguous 
& ``you never know if you are doing it right'' 
& 4 \\

\textit{Too Many Tools} 
& There is not a single tool that fulfills all tasks 
& ``throwing [the data] through a lot of tools'' 
& 2 \\

\midrule
\multicolumn{4}{@{}l}{\textbf{Problems With Tools}} \\

$\hookrightarrow$ \textit{NVivo: Sharing Files} 
& NVivo does not share folders and files seamlessly 
& ``Sharing [\dots] is a disaster in NVivo'' 
& 1 \\

$\hookrightarrow$ \textit{Excel: Adequateness} 
& Excel does not quite do justice to many tasks 
& ``Excel quickly gets too complex for unstructured data'' 
& 3 \\

$\hookrightarrow$ \textit{MAXQDA: Import/Export Formats} 
& MAXQDA imports and exports data incorrectly 
& ``Dislikes: [\dots] export/import'' 
& 2 \\

$\hookrightarrow$ \textit{Audio Transcription: Low Quality} 
& The audio transcriptions are of poor quality and difficult to follow 
& ``Audio transcriptions could be more accurate'' 
& 1 \\

$\hookrightarrow$ \textit{File Management} 
& Keeping track of all files and file versions is hard 
& ``Manual transfer to Excel spreadsheet was tedious'' 
& 1 \\

$\hookrightarrow$ \textit{MAXQDA: Initial Setup and Learning} 
& MAXQDA is complicated and not intuitive to use 
& ``Dislikes: Initial setup was difficult'' 
& 1 \\

$\hookrightarrow$ \textit{MAXQDA: Collaboration} 
& It is not possible to work on a file together in MAXQDA 
& ``you can't work on a project in sync'' 
& 2 \\

$\hookrightarrow$ \textit{MAXQDA: Merging Projects} 
& Difficult and incorrect merging of projects in MAXQDA 
& ``Merging is pretty unintuitive'' 
& 2 \\

$\hookrightarrow$ \textit{MAXQDA: Crashes} 
& MAXQDA crashes during editing 
& ``Crashed often, but did not experience any data loss'' 
& 2 \\

$\hookrightarrow$ \textit{MAXQDA: IRR Computation} 
& MAXQDA does not calculate IRR correctly 
& ``Limited tool support for incorrect or unexplained results where appropriate'' 
& 2 \\

$\hookrightarrow$ \textit{MAXQDA: Too Many/Complex Functions} 
& MAXQDA has too many and too complicated functions 
& ``MAXQDA: Basic use, not using the full feature set, tedious manual process'' 
& 3 \\

\bottomrule
\end{tabular}
\end{table*}

\begin{table*}[t]
\small
\centering
\caption{Codebook for \textit{Willingness} based on our interviews with 15 participants. The codebook includes all codes, along with their descriptions, examples, and the number of participants per code.}
\label{qdallm::tab::codes_willingness}

\setlength{\tabcolsep}{4pt}
\renewcommand{\arraystretch}{1.08}

\begin{tabular}{@{}p{2cm}p{6cm}p{6cm}r@{}}
\toprule
\textbf{Code} & \textbf{Description} & \textbf{Examples} & \textbf{Qty.} \\
\midrule

\textit{Yes} 
& Participant is willing to use AI without conditions 
& ``Willingness to use AI support: Yes'' 
& 2 \\

\textit{Conditional} 
& Participant would use AI, but immediately mentions concerns or conditions 
& ``Yes, but I do not want it to have full control over my data'' 
& 10 \\

\textit{No} 
& Participant is unwilling to use AI in qualitative coding 
& -- 
& 1 \\

\textit{Unclear} 
& Participant did not state willingness 
& ``I want to see how it can complement the research process'' 
& 2 \\

\bottomrule
\end{tabular}
\end{table*}

\begin{table*}[p]
\footnotesize
\centering
\caption{Codebook for \textit{Concerns and Conditions} based on our interviews with 15 participants. The codebook includes all codes structured by high-level concepts, along with their descriptions, examples, and the number of participants per code.}
\label{qdallm::tab::codes_concerns}

\setlength{\tabcolsep}{3.5pt}
\renewcommand{\arraystretch}{1.02}

\begin{tabular}{@{}p{4cm}p{6cm}p{6cm}r@{}}
\toprule
\textbf{Code} & \textbf{Description} & \textbf{Examples} & \textbf{Qty.} \\
\midrule

\multicolumn{4}{@{}l}{\textbf{Scientific Practices}} \\

$\hookrightarrow$ \textbf{Moral / Ethical Concerns} 
& & & \\

\hspace{5mm}$\hookrightarrow$ \textit{Confidentiality} 
& New knowledge should be kept secret 
& ``Confidentiality. Especially when something [\dots] is new'' 
& 1 \\

\hspace{5mm}$\hookrightarrow$ \textit{Sensitive Topics} 
& AI may not deal adequately with sensitive information 
& ``depending on the sensitivity'' 
& 3 \\

\hspace{5mm}$\hookrightarrow$ \textit{Privacy} 
& Concerns that AI may not be able to comply with the privacy policy 
& ``Concerns: privacy, GDPR compliance'' 
& 10 \\

\hspace{5mm}$\hookrightarrow$ \textit{Personal Involvement} 
& Need to be personally involved and know their own data 
& ``Do not make people too dependent on the tools'' 
& 10 \\

\hspace{5mm}$\hookrightarrow$ \textit{Transparency to Participants} 
& Participants should be informed about the use of LLMs 
& ``tell participants that we are analyzing [\dots] with LLM?'' 
& 4 \\

\hspace{5mm}$\hookrightarrow$ \textit{Integrity} 
& AI can threaten the integrity and independence of academic research 
& ``Academic integrity: response can sound good but not be'' 
& 2 \\

$\hookrightarrow$ \textbf{Organizational} 
& & & \\

\hspace{5mm}$\hookrightarrow$ \textit{Implications for ERB} 
& The ERB must be informed of the use of AI 
& ``Implications for ERB, what tools will be used?'' 
& 1 \\

\hspace{5mm}$\hookrightarrow$ \textit{Conference Policies} 
& Paper may be rejected because conference does not allow AI-coding 
& ``the policies \& guidelines of conferences [\dots] around LLMs'' 
& 2 \\

\midrule
\multicolumn{4}{@{}l}{\textbf{Features}} \\

$\hookrightarrow$ \textit{Compatibility} 
& Compatibility with existing tools 
& ``broad compatibility'' 
& 1 \\

$\hookrightarrow$ \textbf{Control/Agency} 
& & & \\

\hspace{5mm}$\hookrightarrow$ \textit{Comparison with Human} 
& Coding should be compared to coding by a human 
& ``If the AI came up with the same results [as a human] multiple times'' 
& 6 \\

\hspace{5mm}$\hookrightarrow$ \textit{Control Codebook Creation} 
& Researcher has full control over codebook content 
& ``So I would have a special look over his shoulder'' 
& 7 \\

\hspace{5mm}$\hookrightarrow$ \textit{Verify Outputs} 
& The researcher must verify the data output by the AI 
& ``look over 20 percent, similar to the human process'' 
& 11 \\

$\hookrightarrow$ \textit{Offline Usage} 
& Offline operation of AI; data is not uploaded to a cloud server 
& ``First pass for anonymization model must run locally'' 
& 2 \\

$\hookrightarrow$ \textit{Collaborative Assistant} 
& Work with AI as a collaborator or interactive assistant 
& ``you work together, i would go through suggestions'' 
& 8 \\

$\hookrightarrow$ \textit{Glitch Free} 
& It should be possible to work with the tool without any errors or problems 
& ``A glitch-free collaboration feature is really what I want'' 
& 1 \\

$\hookrightarrow$ \textit{Changing AI Parameters} 
& Participant wants to be able to tune the AI 
& ``fine-tune things, tinker with different variables'' 
& 7 \\

$\hookrightarrow$ \textit{Explainability} 
& AI has an explanation for its response 
& ``Can they provide justification or explanation?'' 
& 7 \\

$\hookrightarrow$ \textit{Evaluation of AI Performance} 
& Knowing about AI limitations 
& ``I am an observer of how good they are'' 
& 5 \\

\midrule
\multicolumn{4}{@{}l}{\textbf{Meta Concerns}} \\

$\hookrightarrow$ \textit{Learning From Coding} 
& Humans learn by coding, gaining experience over time. AI does not 
& ``Qualitative research is very subjective, [\dots] where does it come from when AI does it?'' 
& 2 \\

$\hookrightarrow$ \textit{Tuning AI Takes Time} 
& Checking and adjusting the AI takes time 
& ``[AI] should provide support and not extra work'' 
& 4 \\

$\hookrightarrow$ \textit{Taking Away Jobs} 
& Concerns about AI replacing human workers 
& ``workplace at risk'' 
& 1 \\

$\hookrightarrow$ \textit{Costs of AI Tool} 
& AI tool may be too expensive, leading to disparities between research institutions 
& ``cost of these tools could also be a concern'' 
& 1 \\

\midrule
\multicolumn{4}{@{}l}{\textbf{AI Output Quality}} \\

$\hookrightarrow$ \textit{Hallucinations} 
& AI cannot provide completely accurate data and has produced incorrect results 
& ``afraid to let it write a whole background section because of wrong citations'' 
& 5 \\

$\hookrightarrow$ \textit{Low Quality} 
& Low level of confidence in AI coding; uncertainty about how to check it 
& ``we give it to the machine, how do we know its [\dots] good?'' 
& 4 \\

$\hookrightarrow$ \textit{Model Will Not Understand} 
& AI does not have the knowledge that humans have to understand 
& ``Qualitative coding is not a straightforward approach'' 
& 3 \\

$\hookrightarrow$ \textit{No Trust in Full AI Control} 
& AI-generated results are in doubt 
& ``Full agency: do not just trust in the results'' 
& 5 \\

$\hookrightarrow$ \textit{Subjective Data Is Hard for AI} 
& It is difficult for AI to interpret and understand subjective statements 
& ``there is no way to rank, e.g., values that are more important than others, at least not for a machine.'' 
& 5 \\

$\hookrightarrow$ \textit{Different Background} 
& We learn from experience, AI does not 
& ``don't think the AI understands what anyone means'' 
& 6 \\

$\hookrightarrow$ \textit{AI Bias} 
& The AI could inadvertently develop biases as it ingests data 
& ``Qualitative research is very subjective, [\dots] where does it come from when AI does it?'' 
& 5 \\

$\hookrightarrow$ \textit{Conversations Unlike With Collaborators} 
& Talking to an AI is not the same as talking to a human, and makes codebook development more difficult 
& ``Translating what you give to a human coder into what you give to an algorithm'' 
& 3 \\

\bottomrule
\end{tabular}
\end{table*}

\clearpage
\section{Examples of Participants' Coding Workflow Drawings}
\label{qdallm::sec::appendix2}

\begin{figure*}[htb]
    \centering
    \includegraphics[width=\linewidth]{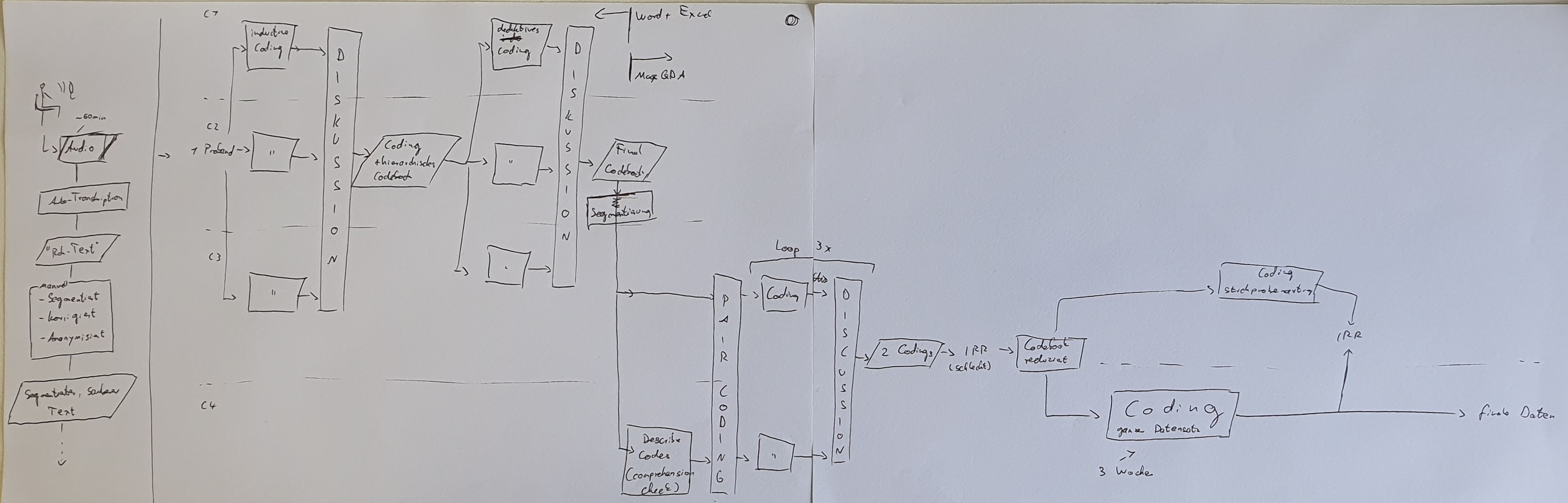}
    
    \includegraphics[width=0.492\linewidth]{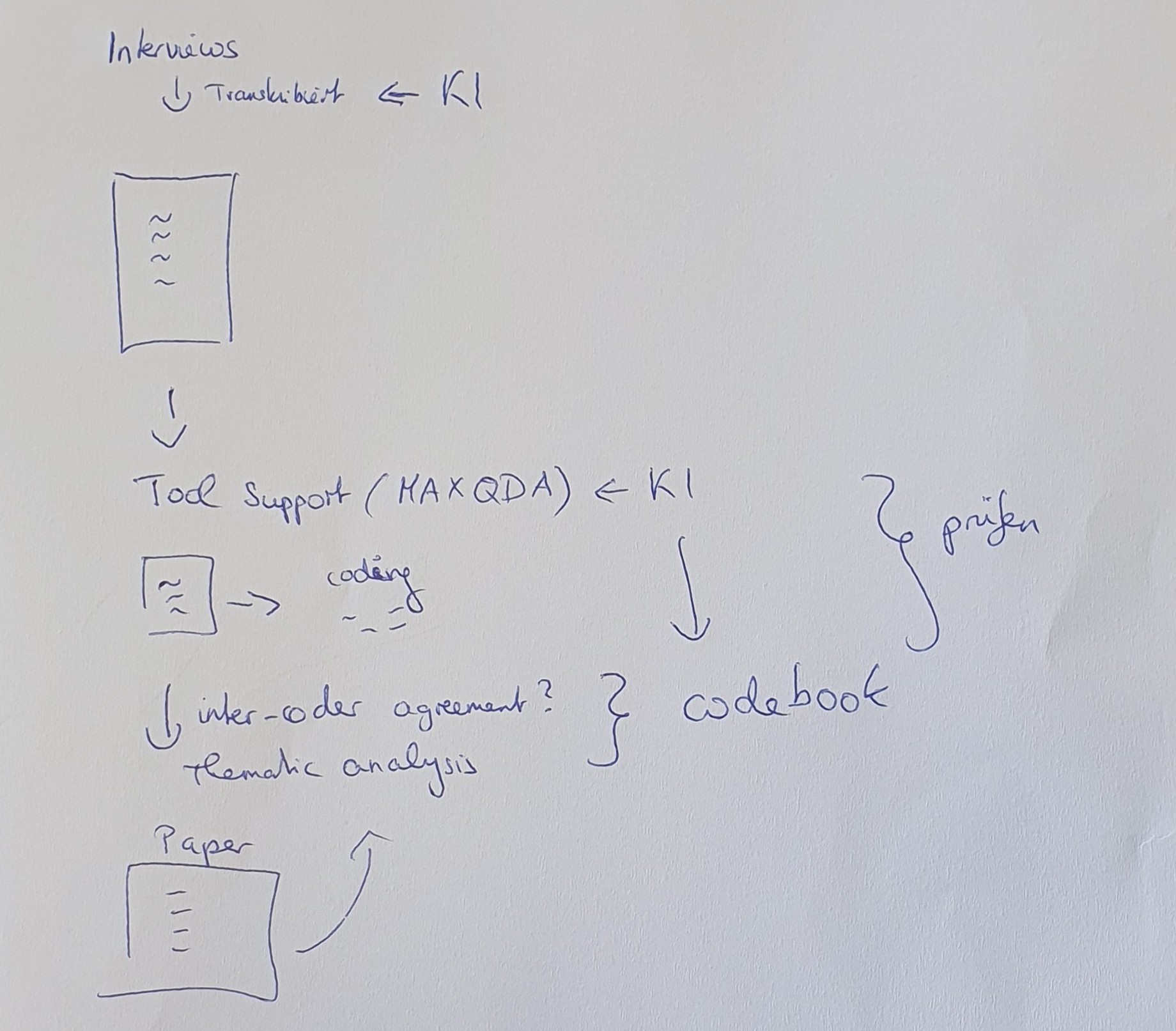}
    \hfill
    \begin{subfigure}{0.49\textwidth}
    \includegraphics[width=\linewidth]{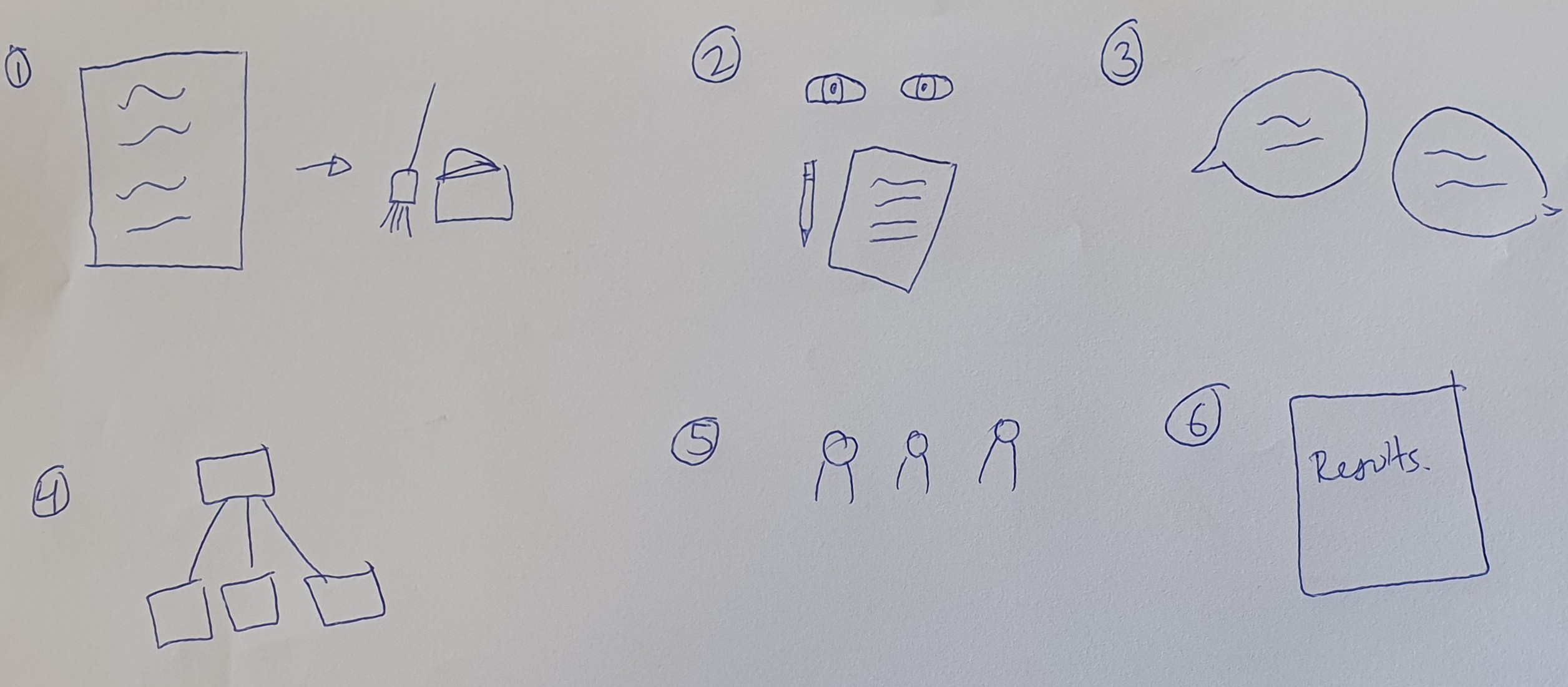}
    \includegraphics[width=\linewidth]{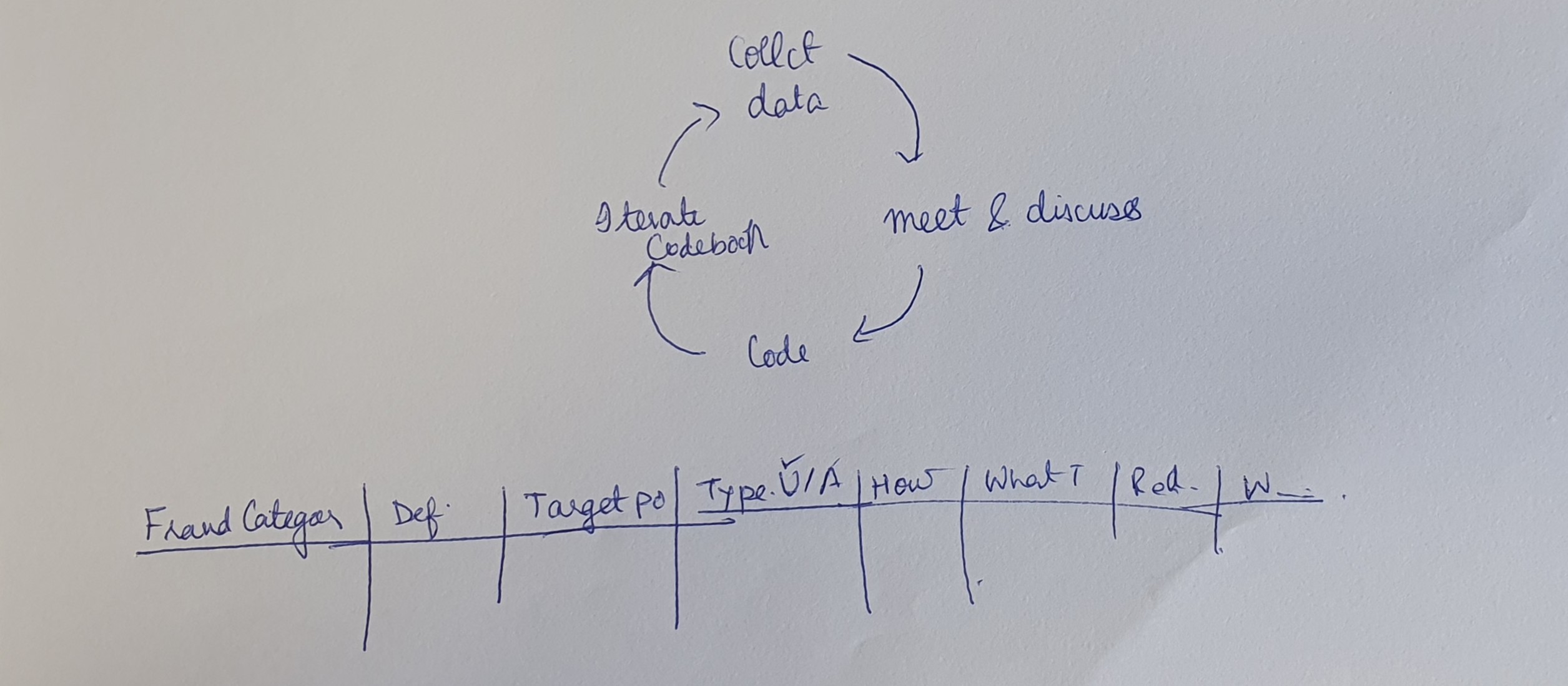}
\end{subfigure}
    
    \caption{Four examples of coding workflow drawings created by our participants during interviews. The drawings varied considerably in the level of detail they provided. Some included in-depth descriptions of specific actions during each step, while others focused more on visuals, providing details of the process only verbally.}
    \label{fig:example-drawings}
    \Description{This figure shows four different photos of workflow drawings created by participants during the interviews. One workflow is very complex, containing many substeps. The second workflow includes only three main steps (Interviews, followed by coding using tool support, and finally, the paper itself). The participant wrote short notes, such as "intercoder agreement" and "thematic analysis," next to the arrows. The third workflow contains only icons and no text. It depicts a six-step process: 1) a piece of paper and a broom, 2) a piece of paper, a pen, and eyes, 3) two speech bubbles, 4) a mind map, 5) three people, and 6) a piece of paper labeled "Results." The last picture is simpler; it contains only a circle consisting of four steps: Collect data, meet and discuss, code, and iterate codebook. Below this, the participant created a small table indicating the columns their typical codebook would have.}
\end{figure*}

\end{document}